\documentclass[superscriptaddress,aps,letterpaper,10pt,twocolumn,floats,showpacs,amsmath,amsfonts,amssymb,prb]{revtex4-2}

\usepackage{graphicx}% Include figure files
\usepackage[colorlinks=true,citecolor=blue]{hyperref}
\usepackage[caption=false]{subfig}
\usepackage{pstricks}
\usepackage{pst-node}
\usepackage{amsmath}
\usepackage{amsbsy}
\usepackage{verbatim}
\usepackage{dsfont}
\usepackage{MnSymbol}
\begin{document}

%\preprint{APS/123-QED}

\title{Bounds on skewness and kurtosis of steady state currents}% Force line breaks with \\

\author{Krzysztof Ptaszy\'{n}ski}
\affiliation{Institute of Molecular Physics, Polish Academy of Sciences, Mariana Smoluchowskiego 17, 60-179 Pozna\'{n}, Poland}
\email{krzysztof.ptaszynski@ifmpan.poznan.pl}
%\author{Massimiliano Esposito}
%\affiliation{Complex Systems and Statistical Mechanics, Physics and Materials Science Research Unit, University of Luxembourg, L-1511 Luxembourg, Luxembourg}

\date{\today}% It is always \today, today,

\begin{abstract}
Current fluctuations are a powerful tool to unravel the underlying physics of the observed transport process. This work discusses some general properties of the third and the fourth current cumulant (skewness and kurtosis) related to dynamics and thermodynamics of a transport setup. Specifically, several distinct bounds on these quantities are either analytically derived or numerically conjectured, which are applicable to: 1) noninteracting fermionic systems, 2) noninteracting bosonic systems, 3) thermally driven classical Markovian systems, 4) unicyclic Markovian networks. Finally, it is demonstrated that violation of the obtained inequalities can provide a broad spectrum of information about the physics of the analyzed system, e.g., enable one to infer the presence of interactions or unitary dynamics, unravel the topology of the Markovian network, or characterize the nature of thermodynamic forces driving the system. In particular, relevant information about the microscopic dynamics can be gained even at equilibrium when the current variance -- a standard measure of current fluctuations -- is determined mostly by the thermal noise.
\end{abstract}

\maketitle

\section{Introduction}
Dynamics at the nanoscale level is inherently stochastic which leads to fluctuations of the observed currents. A notable example is the electronic shot noise resulting from the discrete nature of the electric charge~\cite{blanter2000}. Multiple studies have demonstrated that current fluctuations are not only a nuisance but may reveal important information about physics of the underlying transport process. Due to this fact they have been investigated, both experimentally and theoretically, in the variety of physical contexts, including electronic transport~\cite{bulka2000, belzig2005, sanchez2012, gustavsson2006, ubbelohde2013, hasler2015, nazarov2009, korotkov1994, fricke2007, lau2016, fujisawa2006, gustavsson2007, gustavsson2009, ubbelohde2012, kambly2011}, chemical reactions~\cite{schaewitz2005, kolomeisky2007, moffitt2014, chemla2008, moffitt2010} or optical systems~\cite{carmichael1989, matthiesen2014, kiilerich2014}.

Though current fluctuations are strongly sensitive to details of microscopic dynamics, some of their properties can be characterized by certain physical laws with different ranges of applicability. A notable example is the steady-state fluctuation theorem~\cite{andrieux2009, esposito2009}
\begin{align} \label{flucttheor}
	\frac{P(\sigma)}{P(-\sigma)}=e^{\sigma/k_B},
\end{align} 
which is valid (in the long-time limit) for an arbitrary open quantum system with time-independent parameters; here $P(\sigma)$ is the probability of the entropy production $\sigma$. The other important law is the thermodynamic uncertainty relation bounding the minimum value of current fluctuations, which is applicable to classical Markovian systems with time-independent parameters. This relation reads~\cite{barato2015b, pietzonka2016, gingrich2016, polettini2016, pietzonka2016, pietzonka2017b, horowitz2017, pietzonka2016b, falasco2020}
\begin{align} \label{unc}
	\frac{\llangle j^2 \rrangle}{\llangle j^1 \rrangle^2} \geq \frac{2 k_B}{\dot{\sigma}},
\end{align}
where $\dot{\sigma}$ is the entropy production rate while $\llangle j^1 \rrangle$ and $\llangle j^2 \rrangle$ are the first and the second cumulants of an arbitrary thermodynamic current (the average current and the current variance, respectively); see Sec.~\ref{sec:def} for a detailed definition of current cumulants. The thermodynamic uncertainty relation, in its range of validity, can be used to infer the minimum value of energy dissipation. On the other hand, breaking of Eq.~\eqref{unc} implies violation of the underlying assumption of classical Markovianity due to either a classical underdamped dynamics~\cite{brandner2018, chun2019, fisher2020, pietzonka2022} or quantum effects~\cite{brandner2018, ptaszynski2018, saryal2019, agarwalla2018}. Finally, another important relation states that the Fano factor of the particle current $F=\llangle j_p^2 \rrangle/|\llangle j^1_p \rrangle|$ does not exceed 1 in noninteracting fermionic systems in the high voltage regime~\cite{blanter2000}; therefore, noise enhancement to super-Poissonian values ($F>1$) implies the presence of interactions~\cite{bulka2000, belzig2005}. Additionally, interactions can be also revealed by factorial cumulants of the charge current~\cite{kambly2011}.

While most studies so far focused on the properties of the second current cumulant, namely, the current variance, this paper deals with the third and the fourth cumulant. More specifically, the quantities analyzed are the normalized skewness and kurtosis
\begin{align}
	\mathcal{S}&=\frac{\llangle j^3 \rrangle}{\llangle j^1 \rrangle}, \\
	\mathcal{K}&=\frac{\llangle j^4 \rrangle}{\llangle j^2 \rrangle},
\end{align}
where $\llangle j^n \rrangle$ is $n$th current cumulant; depending on the considered current, they can be either dimensional or dimensionless quantities. The first quantity measures the asymmetry of probability distribution while the second one -- the weight of distribution tails. Skewness and kurtosis have been previously theoretically applied to investigate phenomena such as quantum interference~\cite{wang2007, urban2008, ho2019}, Kondo effect~\cite{komnik2005}, non-Markovian effects~\cite{xue2015}, cotunneling~\cite{emary2009}, Andreev tunneling~\cite{cuevas2003, braggio2011}, spin blockade~\cite{zhang2013}, or detector-induced backaction~\cite{li2013} in nanoelectronic systems. Most notably, in tunnel junctions the third and the first cumulants of the charge current have been found to be directly proportional to each other as $\llangle j_q^3 \rrangle =e^2 \llangle j^1_q \rrangle$, where $e$ is the particle effective charge. In contrast to the current variance, this relation is not affected by the thermal noise which enables one to determine the effective charge even in close-to-equilibrium conditions~\cite{levitov2004}. On the experimental side, though measurement of higher-order current fluctuations is still challenging, it has been already employed to explore the role of intrinsic and environmental contributions to current fluctuations in tunnel junctions~\cite{reulet2003, fevrier2020} and their dynamics under AC driving~\cite{gabelli2009, forgues2013, fevrier2020}, charge multiplication in avalanche diodes~\cite{gabelli2009b}, or crossover from elastic to inelastic transport in short diffusive conductors~\cite{pinsolle2018}. Furthermore, characterization of a full probability distribution of the transmitted charge has been realized in quantum dot systems by means of electron counting methods~\cite{gustavsson2006, fujisawa2006, gustavsson2007, gustavsson2009, ubbelohde2012}.

Only a few studies so far investigated the universal properties of higher current cumulants. Most notably, using the fluctuation theorem~\eqref{flucttheor} universal relations between higher-order current cumulants and nonlinear transport coefficients have been derived~\cite{saito2008}. These relation imply, for example, that the third cumulant $\llangle j^3 \rrangle$ vanishes in time-reversal symmetric systems, while may be finite for a broken time-reversal symmetry~\cite{utsumi2009}. Furthermore, in time-reversal symmetric systems skewness in the linear response regime $\mathcal{S}_\text{lin}$ is equal to the equilibrium kurtosis $\mathcal{K}_\text{eq}$. Both quantities have been further demonstrated to be nonnegative in classical Markovian systems,
\begin{align} \label{eqbound}
	\mathcal{S}_\text{lin}=\mathcal{K}_\text{eq} \geq 0,
\end{align}
which is a direct consequence of the thermodynamic uncertainty relation~\eqref{unc}~\cite{saryal2019}; see Sec.~\ref{subsec:neg} for further details. Additionally, Barato and Seifert~\cite{barato2015} obtained bounds on skewness and kurtosis of waiting times between successive stochastic transitions dependent on the topology of the Markovian network; similar inequalities related to system thermodynamics have been also later conjectured~\cite{wampler2021}.

This article, in Sec.~\ref{sec:bounds}, presents new bounds on skewness and kurtosis applicable to: 1) noninteracting fermionic systems, 2) nonintereacting bosonic systems, 3) thermally driven classical Markovian systems, 4) unicyclic Markovian networks. They are obtained using either analytical derivations or a strong numerical conjecture. Section~\ref{sec:count} presents the exemplary systems in which these bounds can be violated by going beyond their range of applicability. This demonstrates how breaking of the obtained inequalities can be used to infer useful information about the physical system underlying the observed transport process, such as presence of interactions, nature and number of thermodynamics forces driving the system, topology of the Markovian network, or presence of a unitary component of the dynamics. Finally, Sec.~\ref{sec:concl} brings conclusions following from the results.

\section{Definitions} \label{sec:def}
Before presenting the results, let me first define the quantities of interest. The paper will consider fluctuations of a generic stochastic current $j(t)$, for example, charge, heat or particle current. It is useful to define the time integrated current
\begin{align}
	J_t = \int_0^t j(\tau) d\tau.
\end{align}

An important quantity characterizing the current fluctuations is the cumulant generating function
\begin{align}
	G(\lambda,t) = \ln \int_{-\infty}^\infty \rho(J_t) e^{\lambda J_t} d J_t
\end{align}
where $\rho(J_t)$ is the probability density distribution of the integrated current in the moment $t$. It has been demonstrated that in the long time limit the cumulant generating functions grows linearly in time as $G(\lambda,t)=t \chi(\lambda)$ where
\begin{align}
	\chi(\lambda)= \lim_{t \rightarrow \infty} \frac{G(\lambda,t)}{t}
\end{align}
is referred to as the scaled cumulant generating function~\cite{touchette2009}. It can be expressed as a power series
\begin{align}
	\chi (\lambda)=\sum_{n=1}^\infty \frac{\llangle j^n \rrangle \lambda^n}{n!},
\end{align}
where coefficients $\llangle j^n \rrangle$ are known as the scaled cumulants; since the paper deals solely with the steady state properties, they will be referred as cumulants for simplicity. The scaled cumulants can be calculated using $\chi(\lambda)$ as
\begin{align}
	\llangle j^n \rrangle = \left[\frac{\partial^n}{\partial \lambda^n} \chi(\lambda) \right]_{\lambda=0}.
\end{align}

The physical meaning of the scaled cumulants can be revealed through their relation to central moments of the integrated current~\cite{cornish1937}
\begin{align}
	\llangle j^1 \rrangle &= \lim_{t \rightarrow \infty} t^{-1} \langle J_t \rangle, \\
	\llangle j^2 \rrangle &= \lim_{t \rightarrow \infty} t^{-1} \langle \Delta J_t^2 \rangle, \\
	\llangle j^3 \rrangle &= \lim_{t \rightarrow \infty} t^{-1} \langle \Delta J_t^3 \rangle, \\
	\llangle j^4 \rrangle &=\lim_{t \rightarrow \infty} t^{-1} \left( \langle \Delta J_t^4 \rangle -3 \langle \Delta J_t^2 \rangle^2 \right),
\end{align}
where $\Delta J_t=J_t-\langle J_t \rangle$. In particular, the first scaled cumulant is the average current while the second is the current variance.

\section{Bound on skewness and kurtosis} \label{sec:bounds}
I will now present the obtained bounds on skewness and kurtosis. Subsection~\ref{subsec:sum} will summarize the results while the next subsections will provide their justification and deeper discussion. 

\subsection{Summary} \label{subsec:sum}
The obtained bounds read as follows:
\begin{enumerate}
	\item For noninteracting fermionic systems (dimensionless) kurtosis of the particle current $j_p$ obeys the relation
	\begin{align} \label{nonintkurt}
		\mathcal{K}^p &\in \left[-\frac{1}{2},1 \right].
	\end{align}
Here index $p$ refers to the particle current. Additionally, for junctions driven only by a single voltage
	\begin{align} \label{nonintfskew}
	\mathcal{S}^p &\in \left[-\frac{1}{2},1 \right].
\end{align}
    \item For noninteracting bosonic systems kurtosis of the particle current $j_p$ and the heat current $j_h$ is always nonnegative:
\begin{align} \label{kurtbos}
	\mathcal{K}^p,{ }\mathcal{K}^h &\geq 0.
\end{align}
Here index $h$ refers to the heat current. Additionally, for systems driven only by a single thermodynamic force (difference of either bath temperatures or chemical potentials) also skewness is nonnegative:
\begin{align} \label{skewbos}
	\mathcal{S}^p,{ }\mathcal{S}^h &\geq 0.
	\end{align}
The same relations apply to heat transport in classical harmonic systems, which are a classical limit of noninteracting bosonic systems.
\item In classical Markovian systems driven only by temperature differences kurtosis of the heat current is always nonnegative
\begin{align} \label{kurtmth}
	\mathcal{K}^h \geq 0.
\end{align}
Additionally, for systems driven by a single temperature difference also skewness of the heat current is nonnegative
\begin{align} \label{skewmth}
 \mathcal{S}^h \geq 0.
\end{align}
\item For unicyclic Markovian networks skewness and kurtosis of the winding number (i.e., number of rotations around the cycle) obey the relations
\begin{align} \label{skewuni}
	\mathcal{S} &\in \left[-\frac{1}{16}, 1 \right], \\ \label{kurtuni}
	\mathcal{K} &\in \left[-\frac{\sqrt{5}+1}{10},1 \right], \\ \label{difuni}
		\mathcal{K}-\mathcal{S} &\in \left[-0.15, 0.465 \right], \\ \label{sumuni}
\mathcal{K}+\mathcal{S} &\in \left[-\frac{8}{27},2 \right], \\ \label{produni}
\mathcal{K} \times \mathcal{S} &\in \left[-0.054, 1\right].	
\end{align}
Here the values denoted with decimal numerals are approximate.
\end{enumerate}

\subsection{Noninteracting fermionic systems} \label{subsec:ferm}
This section provides a justification of Eqs.~\eqref{nonintkurt}--\eqref{nonintfskew}. In particular, in Sec.~\ref{sec:timerev} these inequalities will be analytically derived for time-reversal symmetric systems, whereas in Sec.~\ref{sec:btimerev} a numerical verification of Eq.~\eqref{nonintkurt} for systems with a broken time-reversal symmetry will be presented.

\subsubsection{Time-reversal symmetric case} \label{sec:timerev}
Here I consider transport in a generic multiterminal fermionic junction  consisting of a scattering system (for example, quantum dot) coupled to $L$ baths (or leads) $\alpha$ with temperatures $T_\alpha$ [inverse temperatures $\beta_\alpha=1/(k_B T_\alpha)$] and chemical potential $\mu_\alpha$. Such systems are commonly analyzed in the context of mesoscopic electronic transport~\cite{nazarov2009}. When the interelectron interactions can be neglected, the whole system (including scattering region and the baths) can be described by a quadratic Hamiltonian of a general form
\begin{align}
	H_{NF}= \sum_{ij} \left( t_{ij} d_i^\dagger d_j + \text{h.c.} \right),
\end{align}
where $d_i^\dagger$ ($d_i$) is the fermionic creation (annihilation) operator.

The section will focus on scaled cumulants of the particle current $j_p$, which is defined as the number of particles (e.g. electrons) transmitted per unit of time. All scaled cumulants of the particle current $\llangle j^n_p \rrangle$ have a dimension of 1/s, which makes $\mathcal{S}$ and $\mathcal{K}$ dimensionless. By definition, they are directly related to the cumulants of the charge current $j_q$ as $\llangle j_q^n \rrangle=e^n \llangle j_p^n \rrangle$, where $e$ is the particle charge. In the noninteracting case cumulants of the particle current flowing to the bath $\alpha$, denoted as $\llangle j_{p,\alpha}^n \rrangle$, are sums of cumulants of currents flowing from different baths $\gamma \neq \alpha$, denoted as $\llangle j_{p,\gamma \rightarrow \alpha}^n \rrangle$:
\begin{align} \label{sumcum}
	\llangle j_{p,\alpha}^n \rrangle=\sum_{\gamma \neq \alpha} \llangle j_{p,\gamma \rightarrow \alpha}^n \rrangle.
\end{align}
This is no longer true in interacting systems due to presence of electron correlations. 

For time-reversal symmetric systems cumulants can be calculated using the equation
\begin{align} \label{fermcum}
	\llangle j_{p,\gamma \rightarrow \alpha}^n \rrangle = \left[ \frac{\partial^n}{\partial \lambda^n} \chi^p_{\alpha \gamma}(\lambda) \right]_{\lambda=0},
\end{align}
with the scaled cumulant generating function $\chi^p_{\alpha \gamma}(\lambda)$ given by the Levitov-Lesovik formula~\cite{levitov1993}
\begin{align} \label{levitov} \nonumber 
	\chi^p_{\alpha \gamma}(\lambda)=\int_{-\infty}^{\infty} \frac{d \omega}{2\pi} & \ln \left \{1+\mathcal{T}_{\alpha \gamma}(\omega) \left [\left (e^{\lambda}-1 \right) f_\gamma (\omega) g_\alpha (\omega) \right. \right. \\ 
	& \left. \left.  +\left (e^{-\lambda}-1 \right) f_\alpha (\omega) g_\gamma (\omega) \right] \right \},
\end{align}
where $\mathcal{T}_{\alpha \gamma}(\omega)$ is the transmission function taking values within the range $[0,1]$, $f_\alpha(\omega)=1/\{1+\exp[\beta_\alpha(\omega-\mu_\alpha)]\}$ is the Fermi distribution function of the bath $\alpha$ and $g_\alpha(\omega)=1-f_\alpha(\omega)$. Here and from hereon $\hbar=1$ is taken. For time-reversal symmetric systems the transmission function is invariant under the index exchange: $\mathcal{T}_{\alpha \gamma}(\omega) = \mathcal{T}_{ \gamma \alpha}(\omega)$. Using Eqs.~\eqref{fermcum} and~\eqref{levitov} one gets
\begin{align} \label{fkern}
	\llangle j_{p,\gamma \rightarrow \alpha}^n \rrangle = \int_{-\infty}^{\infty} \frac{d \omega}{2\pi} C_{\alpha \gamma,n}(\omega),
\end{align}
where $C_{\alpha \gamma,n}(\omega)$ are functions of $\mathcal{T}_{\alpha \gamma}(\omega)$, $f_\alpha(\omega)$ and $f_\gamma(\omega)$. 

Let me now express first four functions $C_{\alpha \gamma,n}(\omega)$ using a simplified notation $\mathcal{T}_{\alpha \gamma}(\omega)=\mathcal{T}_{\alpha \gamma}$ and $f_\alpha(\omega)=f_\alpha$:
\begin{align} \label{cfun1}
&	C_{\alpha \gamma,1}(\omega) =\mathcal{T}_{\alpha \gamma} \left(f_\gamma-f_\alpha \right), \\
&	C_{\alpha \gamma,2}(\omega) =\mathcal{T}_{\alpha \gamma} \left( f_\alpha+f_\gamma-2 f_\alpha f_\gamma \right) + \mathcal{T}^2_{\alpha \gamma} \left( f_\alpha-f_\gamma \right)^2, \\
& C_{\alpha \gamma,3}(\omega) =\mathcal{T}_{\alpha \gamma} \left(f_\gamma-f_\alpha \right)\\ \nonumber
& - 3 \mathcal{T}^2_{\alpha \gamma} \left( f_\gamma-f_\alpha \right) \left( f_\alpha+f_\gamma-2 f_\alpha f_\gamma \right) +2 \mathcal{T}^3_{\alpha \gamma} \left( f_\gamma-f_\alpha \right)^3, \\ \label{cfun4}
& C_{\alpha \gamma,4}(\omega) = \mathcal{T}_{\alpha \gamma} \left( f_\alpha+f_\gamma-2 f_\alpha f_\gamma \right) \\ \nonumber
&-\mathcal{T}_{\alpha \gamma}^2 \left[4 \left(f_\gamma-f_\alpha \right)^2+3 \left( f_\alpha+f_\gamma-2 f_\alpha f_\gamma \right)^2 \right] \\ \nonumber
& +12 \mathcal{T}_{\alpha \gamma}^3 \left(f_\gamma-f_\alpha \right)^2 \left( f_\alpha+f_\gamma-2 f_\alpha f_\gamma \right) \\ \nonumber 
&-6 \mathcal{T}^4_{\alpha \gamma} \left( f_\alpha-f_\gamma \right)^4.
\end{align}

Bound~\eqref{nonintkurt} will be now derived using a general inequality bounding the ratio of sums of two sequences,
\begin{align} \label{sumseq}
	\min_i \frac{a_i}{b_i} \leq \frac{\sum_i a_i}{\sum_i b_i} \leq \max_i \frac{a_i}{b_i},
\end{align}
which is valid when all coefficients $b_i$ are of the same sign. Let me denote
\begin{align}
\mathcal{K}^p_\alpha &= \frac{\llangle j_{p,\alpha}^4 \rrangle}{\llangle j_{p,\alpha}^2 \rrangle}, \\
\mathcal{K}^p_{\alpha \gamma,\omega} &=\frac{C_{\alpha \gamma,4}(\omega)}{C_{\alpha \gamma,2}(\omega)}.
\end{align}
Using Eqs.~\eqref{sumcum}, \eqref{fkern}, \eqref{sumseq} and the inequality $C_{\alpha \gamma,2}(\omega) \geq 0$, which can be easily verified, one gets
\begin{align}
	\min(\mathcal{K}^p_{\alpha \gamma,\omega}) \leq \mathcal{K}^p_\alpha \leq \max(\mathcal{K}^p_{\alpha \gamma,\omega}).
\end{align}
One further finds
\begin{align}
	\begin{cases}
		\min(\mathcal{K}^p_{\alpha \gamma,\omega})=-\frac{1}{2} & \text{for} \quad \mathcal{T}_{\alpha \gamma}=1,{ }f_\alpha=f_\gamma=\frac{1}{2}, \\
		\max(\mathcal{K}^p_{\alpha \gamma,\omega})=1 & \text{for} \quad \mathcal{T}_{\alpha \gamma} \rightarrow 0,
	\end{cases}
\end{align}
which implies $-1/2 \leq \mathcal{K}_\alpha^p \leq 1$ and thus proves Eq.~\eqref{nonintkurt}.

The analogous expression
\begin{align} \label{skewminmax}
	\min(\mathcal{S}^p_{\alpha \gamma,\omega}) \leq \mathcal{S}^p_\alpha \leq \max(\mathcal{S}^p_{\alpha \gamma,\omega})
\end{align}
with
\begin{align}
	\mathcal{S}^p_\alpha &= \frac{\llangle j_{p,\alpha}^3 \rrangle}{\llangle j_{p,\alpha}^1 \rrangle}, \\
	\mathcal{S}^p_{\alpha \gamma,\omega} &=\frac{C_{\alpha \gamma,3}(\omega)}{C_{\alpha \gamma,1}(\omega)},
\end{align}
is, in general, no longer true since the sign of $C_{\alpha \gamma,1}(\omega)$ can be different depending on $\omega$ and $\gamma$. It is valid, however, in a two terminal setup when both leads $\alpha$ and $\gamma$ have the same temperature $T_\alpha=T_\gamma=T$. The sign of $C_{\alpha \gamma,1}(\omega)$ is then independent of $\omega$,
\begin{align}
	\text{sgn} \left[ C_{\alpha \gamma,1}(\omega) \right]= \text{sgn} (\mu_\gamma-\mu_\alpha),
\end{align}
and thus Eq.~\eqref{sumseq} can be applied to obtain Eq.~\eqref{nonintfskew}.

Let us now discuss how Eq.~\eqref{skewminmax}, and thus Eq.~\eqref{nonintfskew}, can be broken beyond its range of validity. First, in a three-terminal setup with leads $\alpha$, $\gamma$, $\delta$ one can tune the chemical potentials to get $\llangle j_{p,\gamma \rightarrow \alpha}^1 \rrangle \approx - \llangle j_{p,\delta \rightarrow \alpha}^1 \rrangle$, such that $\llangle j_{p,\alpha}^1 \rrangle$ is equal or close to 0 while $\llangle j_{p,\alpha}^3 \rrangle$ remains finite. Therefore, $\mathcal{S}^p_\alpha$ can take an arbitrary value from $-\infty$ to $\infty$. Second, the inequality~\eqref{skewminmax} can be broken even in a two-terminal junction when temperatures of the leads are different. In particular, for equal chemical potentials of both leads ($\mu_\alpha=\mu_\gamma=\mu$) the sign of $C_{\alpha \gamma,1}(\omega)$ is a step function of $\omega$:
\begin{align}
	\text{sgn} \left[ C_{\alpha \gamma,1}(\omega) \right]= \text{sgn} \left ( \frac{\omega-\mu}{T_\gamma-T_\alpha} \right).
\end{align}
One can then choose $\mu$ in such a way that $\llangle j_{p,\alpha}^1 \rrangle \approx 0$, since the energy-resolved currents for different $\omega$ compensate, while $\llangle j_{p,\alpha}^3 \rrangle$ remains finite. This again results in a diverging $\mathcal{S}^p_\alpha$.

\subsubsection{Time-reversal asymmetric case} \label{sec:btimerev}
Let us now turn our attention to the case when the time-reversal symmetry is broken (e.g., due to magnetic field). The current fluctuations can be then described by the scaled cumulant generating function $\chi^p(\pmb{\lambda})$, where $\pmb{\lambda}=(\lambda_1,...,\lambda_L)^T$ is the vector of counting fields associated with different baths $\alpha$. It is given by the equation~\cite{levitov1993}
\begin{align}
	\chi^p(\pmb{\lambda})=\int_{-\infty}^\infty \frac{d \omega}{2 \pi} \chi^p_\omega(\pmb{\lambda}),
\end{align}
where
\begin{align}
	\chi^p_\omega(\pmb{\lambda})=\ln \det (\mathds{1}_L -F+F S^\dagger \tilde{S}).
\end{align}
Here $\mathds{1}_L$ is the $L \times L$ identity matrix, $F=\text{diag}(f_1,...,f_L)$ is the diagonal matrix of Fermi distributions and $S$ is the scattering matrix being a $L \times L$ unitary matrix (which becomes Hermitian in the time-reversal symmetric case). Finally, $\tilde{S}$ is the counting-field-dependent scattering matrix with elements
\begin{align}
	\tilde{S}_{\alpha \gamma}=S_{\alpha \gamma} e^{\lambda_\alpha-\lambda_\gamma}.
\end{align}
Cumulants of the particle current to the bath $\alpha$ can be calculated as
\begin{align}
\llangle j_{p,\alpha}^n \rrangle = \int_{-\infty}^\infty \frac{d\omega}{2 \pi} C_{\alpha,n}(\omega),
\end{align}
where
\begin{align}
C_{\alpha,n}(\omega) = \left[ \frac{\partial^n}{\partial \lambda_\alpha^n} \chi_\omega^p(\pmb{\lambda}) \right]_{\pmb{\lambda}=\mathbf{0}}
\end{align}
%
%%%%%%%%%%%%%%%%%%%%%%%%%%%%%%%%%%%%%%%%%%%%%%%%%%%%%%%%%%%%
\begin{figure}
	\centering
	\includegraphics[width=0.90\linewidth]{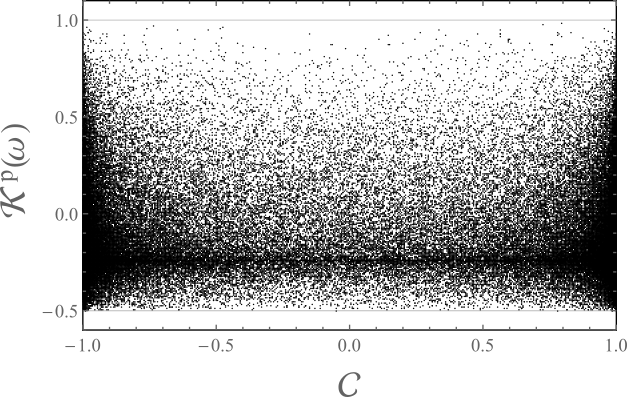}		
	\caption{The function $\mathcal{K}^p(\omega)$ defined in Eq.~\eqref{freqdepkurt} for 100,000 random three-terminal junctions as a function of the circulation coefficient $\mathcal{C}$.}
	\label{fig:randscat}
\end{figure}
%%%%%%%%%%%%%%%%%%%%%%%%%%%%%%%%%%%%%%%%%%%%%%%%%%%%%%%%%%%%
%
with $\mathbf{0}=(0,...,0)^T$.  As in the previous paragraph, to prove Eq.~\eqref{nonintkurt} it is sufficient to show that
\begin{align} \label{freqdepkurt}
	\mathcal{K}^p(\omega)=\frac{C_{\alpha,4}(\omega)}{C_{\alpha,2}(\omega)} \in \left[-\frac{1}{2},1 \right].
\end{align}
This has been numerically verified by simulating more than 100,000 random scattering systems with number of baths $L \leq 6$. In particular, the random scattering matrices have been generated as $S=\exp(iH)$ where $H$ is a random Hermitian matrix. The matrices $H$ have been further generated as $H=G+G^\dagger$, where $\text{Re}(G_{ij})$ and $\text{Im}(G_{ij})$ and random number taken from the uniform distribution over the interval $[0,1]$. The Fermi distribution functions entering the vector $F$ were also taken randomly from the interval $[0,1]$. For illustration, the simulation results for 100,000 random three-terminal junctions have been presented in Fig.~\ref{fig:randscat}. They have been plotted as a function of the circulation coefficient
\begin{align}
 \mathcal{C} = \frac{|S_{12}|^2 |S_{23}|^2 |S_{31}|^2-|S_{21}|^2 |S_{32}|^2 |S_{13}|^2}{|S_{12}|^2 |S_{23}|^2 |S_{31}|^2+|S_{21}|^2 |S_{32}|^2 |S_{13}|^2}
\end{align}
characterizing the asymmetry of the transport setup; it takes value 0 for the time-reversal symmetric case while -1 or 1 for a maximum asymmetry. As one can observe, neither validity or tightness of the bound~\eqref{nonintkurt} depends on the asymmetry: kurtosis takes values within the full range $[-1/2,1]$ for an arbitrary value of $\mathcal{C}$.

\subsection{Noninteracting bosonic systems}
Let us now consider systems of noninteracting bosons described by quadratic Hamiltonians of a general form
\begin{align}
	H_{NB} = \sum_{ij} \left( t_{ij} b_i^\dagger b_j + u_{ij} b_i^\dagger b_j^\dagger + \text{h.c.} \right),
\end{align}
where $b_i^\dagger$ ($b_i$) is the bosonic creation (annihilation) operator. Such models are most commonly applied to describe heat transport in harmonic junctions~\cite{saito2007, saryal2019}, though other types of physical setups, such as junctions of reservoirs of bosonic cold atoms (referred to as the ``atomtronic'' junctions)~\cite{gutman2012}, have been also investigated. The scaled cumulant generating function of the heat current from the bath $\gamma$ to $\alpha$ is now given by the formula~\cite{saito2007, gaspard2015}
\begin{align} \label{levitovbos} 
&	\chi^h_{\alpha}(\lambda) =
	 -\sum_{\gamma \neq \alpha} \int_{0}^{\infty} \frac{d \omega}{2\pi} \times  \\ \nonumber
	& \ln \left \{1-\mathcal{T}_{\alpha \gamma}(\omega) \left [\left (e^{\lambda (\omega-\mu_\alpha)}-1 \right)  \nu_\alpha (\omega) n_\gamma (\omega) \right. \right. \\ \nonumber
	& \left. \left.  +\left (e^{-\lambda (\omega-\mu_\alpha)}-1 \right) n_\alpha (\omega) \nu_\gamma (\omega) \right] \right \},
\end{align}
where $n_\alpha(\omega)=1/\{\exp[\beta_\alpha(\omega-\mu_\alpha)]-1\}$ is the Bose-Einstein distribution of the bath $\alpha$ and $\nu_\alpha(\omega)={n_\alpha(\omega)+1}$; note that for noninteracting bosons the chemical potentials $\mu_\alpha$ are always nonpositive. The scaled cumulant generating function for the particle current is obtained by replacing $\lambda (\omega-\mu_\alpha)$ with $\lambda$. Analogously to the fermionic systems, cumulants of the particle current $\llangle j_{p,\alpha}^n \rrangle$ and the heat current $\llangle j_{h,\alpha}^n \rrangle$ can be calculated using equations
\begin{align} \label{bkern}
	\llangle j_{p,\alpha}^n \rrangle &= \sum_\alpha \int_{0}^{\infty} \frac{d \omega}{2\pi} B_{\alpha \gamma,n}(\omega), \\
	\llangle j_{h,\alpha}^n \rrangle &= \sum_\alpha \int_{0}^{\infty} \frac{d \omega}{2\pi} (\omega-\mu_\alpha)^n B_{\alpha \gamma,n}(\omega),
\end{align}
where $B_{\alpha \gamma,n}(\omega)$ are functions of $\mathcal{T}_{\alpha \gamma}(\omega)=\mathcal{T}_{\alpha \gamma}$, $n_\alpha(\omega)=n_\alpha$ and $n_\gamma(\omega)=n_\gamma$. The first four functions $B_{\alpha \gamma,n}(\omega)$ read as
\begin{align} \label{bfun1}
	&	B_{\alpha \gamma,1}(\omega) =\mathcal{T}_{\alpha \gamma} \left(n_\gamma-n_\alpha \right), \\
	&	B_{\alpha \gamma,2}(\omega) =\mathcal{T}_{\alpha \gamma} \left( n_\alpha+n_\gamma+2 n_\alpha n_\gamma \right) + \mathcal{T}^2_{\alpha \gamma} \left( n_\alpha-n_\gamma \right)^2, \\
	& B_{\alpha \gamma,3}(\omega) =\mathcal{T}_{\alpha \gamma} \left(n_\gamma-n_\alpha \right)\\ \nonumber
	& + 3 \mathcal{T}^2_{\alpha \gamma} \left( n_\gamma-n_\alpha \right) \left( n_\alpha+n_\gamma+2 n_\alpha n_\gamma \right) \\ \nonumber &+2 \mathcal{T}^3_{\alpha \gamma} \left( n_\gamma-n_\alpha \right)^3, \\ \label{bfun4}
	& B_{\alpha \gamma,4}(\omega) = \mathcal{T}_{\alpha \gamma} \left( n_\alpha+n_\gamma+2 n_\alpha n_\gamma \right) \\ \nonumber
	&+\mathcal{T}_{\alpha \gamma}^2 \left[4 \left(n_\gamma-n_\alpha \right)^2+3 \left( n_\alpha+n_\gamma+2n_\alpha n_\gamma \right)^2 \right] \\ \nonumber
	& +12 \mathcal{T}_{\alpha \gamma}^3 \left(n_\gamma-n_\alpha \right)^2 \left( n_\alpha+n_\gamma+2 n_\alpha n_\gamma \right) \\ \nonumber 
	&+6 \mathcal{T}^4_{\alpha \gamma} \left( n_\alpha-n_\gamma \right)^4.
\end{align}
Using expressions above one can easily verify that
\begin{itemize}
	\item $B_{\alpha \gamma,1}(\omega)$ and $B_{\alpha \gamma,3}(\omega)$ are of the same sign,
	\item $B_{\alpha \gamma,2}(\omega)$ and $B_{\alpha \gamma,4}(\omega)$ are nonnegative.
\end{itemize}
Following reasoning presented for the fermionic case, this proves Eq.~\eqref{kurtbos} for the generic case and Eq.~\eqref{skewbos} when $B_{\alpha \gamma,1}(\omega)$ is of the same sign independent of $\gamma$ and $\omega$; the latter holds when the system is driven only by a single thermodynamic force, i.e., for a two-terminal junction with either $T_\alpha=T_\gamma$ or $\mu_\alpha=\mu_\gamma$. 

Bound~\eqref{kurtbos} has been also numerically verified for time-reversal asymmetric systems in a way analogous to described in Sec.~\ref{sec:btimerev} for fermionic systems. In the bosonic case the scaled cumulant generating function of the heat current reads~\cite{gaspard2015}
\begin{align}
	\chi^h(\pmb{\lambda})=\int_{0}^\infty \frac{d \omega}{2 \pi} \chi^h_\omega(\pmb{\lambda}),
\end{align}
where
\begin{align}
	\chi^h_\omega(\pmb{\lambda})=\ln \det (\mathds{1}_L +\mathcal{N}-\mathcal{N} S^\dagger \tilde{S}).
\end{align}
Here $\mathcal{N}=\text{diag}(n_1,...,n_L)$ is the diagonal matrix of the Bose-Einstein distribution and elements of the counting-field-dependent scattering matrix take a form
\begin{align}
	\tilde{S}_{\alpha \gamma}=S_{\alpha \gamma} e^{\lambda_\alpha(\omega-\mu_\alpha)-\lambda_\gamma (\omega-\mu_\gamma)}.
\end{align}

Finally, let us here discuss a qualitative difference between fluctuations of fermionic and bosonic currents. As one may note, Eq.~\eqref{levitovbos} differs from the fermionic Levitov-Lesovik formula \eqref{levitov} by the presence of a minus sign before the whole expression and $\mathcal{T}_{\alpha \gamma}(\omega)$; this is related to different statistical properties of fermions and bosons (particle antibunching and bunching for fermions and bosons, respectively). Accordingly, also expressions for current cumulants have a similar form, but differ by signs [compare Eqs.~\eqref{cfun1}-\eqref{cfun4} and \eqref{bfun1}-\eqref{bfun4}]. As a consequence, kurtosis of the fermionic particle current is confined to a relatively narrow range, while for bosonic systems it may arbitrary nonnegative values.

\subsection{Thermally-driven Markovian systems} \label{subsec:thermmark}
In the next step I will consider heat transport in thermally driven systems evolving according to a classical master equation describing the stochastic transitions between $N$ discrete states of the system. Such models are commonly used in the variety of physical contexts, including electronic transport~\cite{bulka2000, belzig2005, braggio2011, fichetti1998, sanchez2012} or chemical reactions~\cite{barato2015, barato2015b, schaewitz2005, kolomeisky2007}. Let $p_i$ be the probability of the system being in state $i$. The dynamics of the population vector $\mathbf{p}=(p_1,...,p_N)^T$ is given by the equation
\begin{align}
	\dot{\mathbf{p}}=W \mathbf{p}.
\end{align}
Here $W$ is the rate matrix with elements
\begin{align}
	\begin{cases}
	W_{ij}=k_{ij} & \text{for} \quad i \neq j, \\
	W_{ij} = -\sum_{j \neq i} k_{ij} & \text{for} \quad i= j,
\end{cases}
\end{align}
where $k_{ij}$ is the transition rate from the state $j$ to $i$. The transition rates can be expressed as a sum of contributions associated with different baths
\begin{align}
	k_{ij}=\sum_\alpha k_{ij}^\alpha
\end{align}
which obey the detailed balance condition~\cite{seifert2012}
\begin{align} \label{detbal}
	\frac{k_{ij}^\alpha}{k_{ji}^\alpha}=e^{-\beta_\alpha (E_i-E_j)},
\end{align}
where $E_i$ is the energy of state $i$.

To calculate cumulants of the heat current flowing to the bath $\alpha$ one defines the counting-field-dependent rate matrix with elements expressed as~\cite{sanchez2012}
\begin{align}
	\begin{cases}
		[W^h(\pmb{\lambda})]_{ij}=\sum_\alpha k_{ij}^\alpha e^{-\lambda_\alpha (E_i-E_j)} & \text{for} \quad i \neq j, \\
		[W^h(\pmb{\lambda})]_{ij} = -\sum_{j \neq i} k_{ij} & \text{for} \quad i= j,
	\end{cases}
\end{align}
where, as in Sec.~\ref{sec:btimerev}, $\pmb{\lambda}$ is the vector of counting fields. The cumulant generating function $\chi^h(\pmb{\lambda})$ is then equal to the dominant eigenvalue of $W^h(\pmb{\lambda})$~\cite{touchette2009}; however, its analytic calculation is usually not possible for $N >4$. Fortunately, the current cumulants can be determined without direct calculation of $\chi^h(\pmb{\lambda})$ by using a procedure proposed in Refs.~\cite{bruderer2014, wachtel2015}. Within this approach to obtain first $M$ cumulants one writes $M$ equations
\begin{align} \label{calcum1}
	\left \{ \frac{\partial^n}{\partial \lambda_\alpha^n} \det[\mathds{1}\chi^h(\pmb{\lambda})-W^h(\pmb{\lambda})] \right \}_{\pmb{\lambda}=\mathbf{0}} =0,
\end{align}
where $n=1,...,M$ and $\mathds{1}$ is the identity matrix. Such equations are trivially valid since $\chi^h(\pmb{\lambda})$ is an eigenvalue of $W^h(\pmb{\lambda})$ and thus ${\det[\mathds{1} \chi^h(\pmb{\lambda})-W^h(\pmb{\lambda})]=0}$. Upon substituting
\begin{align}
\left[ \frac{\partial^n}{\partial \lambda_\alpha^n} \chi^h(\pmb{\lambda}) \right]_{\pmb{\lambda}=\mathbf{0}} & \rightarrow \llangle j^n_{h,\alpha} \rrangle, \\ \label{calcum3}
\chi^h(\pmb{0})  & \rightarrow  0,
\end{align}
one obtains an easily solvable system of $M$ linear equations with $M$ variables $\llangle j^n_{h,\alpha} \rrangle$.

%
%%%%%%%%%%%%%%%%%%%%%%%%%%%%%%%%%%%%%%%%%%%%%%%%%%%%%%%%%%%%
\begin{figure}
	\centering
	\includegraphics[width=0.90\linewidth]{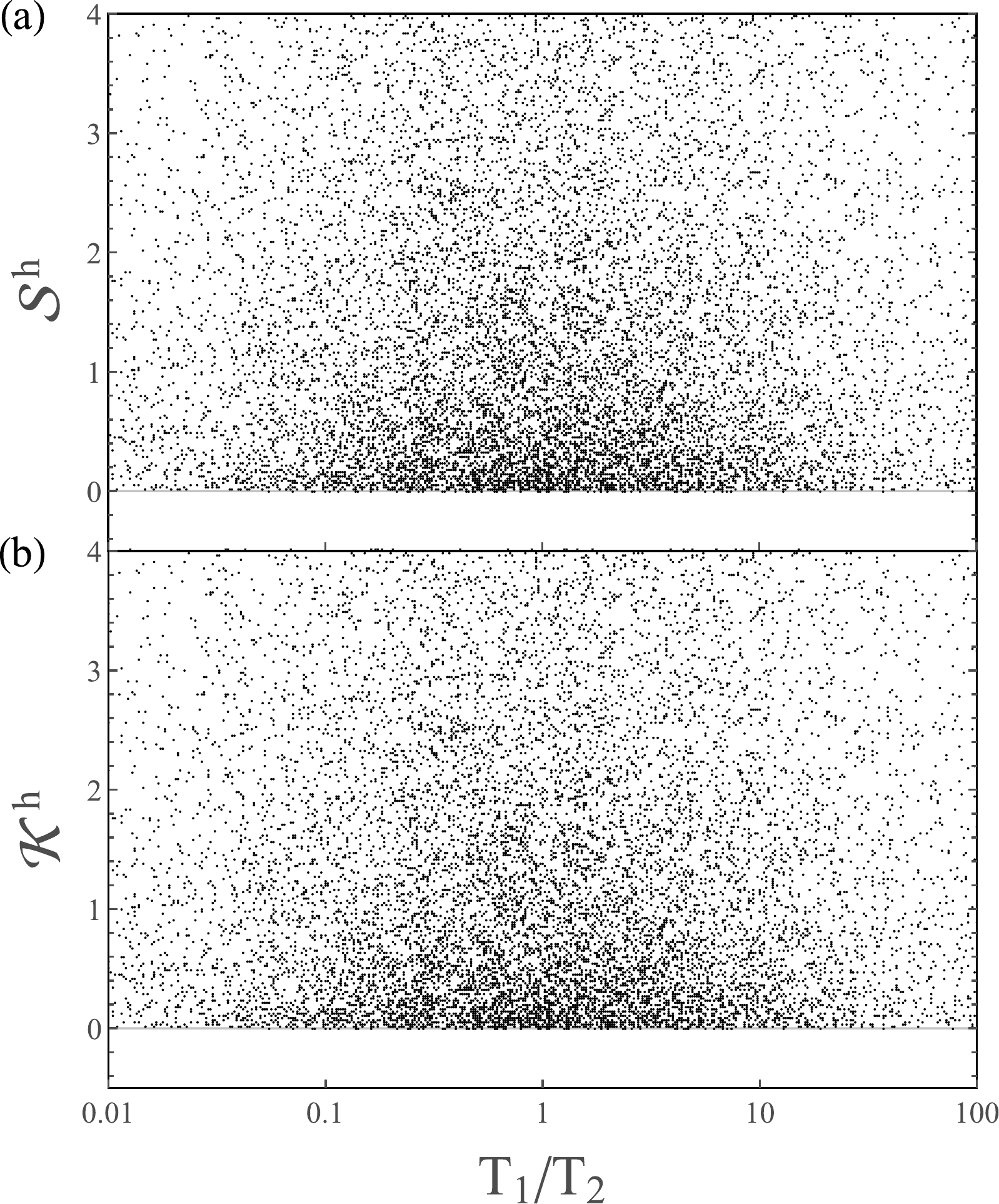}		
	\caption{Skewness and kurtosis of the heat current for 30,000 random four-state Markovian systems attached to two thermal baths as a function of the temperature ratio $T_1/T_2$.}
	\label{fig:markth}
\end{figure}
%%%%%%%%%%%%%%%%%%%%%%%%%%%%%%%%%%%%%%%%%%%%%%%%%%%%%%%%%%%%
%
Bounds~\eqref{kurtmth} and~\eqref{skewmth} have been verified by calculating the heat current cumulants for more than 30,000 random Markovian networks with number of states $N \leq 6$ and number of baths $L \leq 4$. Such networks were generated by choosing random energies $E_i$, temperatures $T_\alpha$ and rates $k_{ij}^\alpha$ for $i>j$; the rates $k_{ij}^\alpha$ for $i<j$ have been then determined using Eq.~\eqref{detbal}; more precisely, the transition rates have been taken from the interval $[0,1]$, while energies and temperatures have been generated as $E_i=e_i/(1-e_i)$ and $T_\alpha=t_\alpha/(1-t_\alpha)$, with $e_i$ and $t_\alpha$ taken from the interval $[0,1]$. For illustration, Fig.~\ref{fig:markth} shows the values of skewness and kurtosis for randomly generated four-state networks attached to two thermal baths as a function of the temperature ratio $T_1/T_2$; as one can observe, they can take arbitrary nonnegative value independent of the temperature ratio.

Bound~\eqref{skewmth} can be broken in the presence of more than a single temperature difference, for example, in a three-terminal system. Furthermore, as will be shown in Sec.~\ref{subsec:dcb}, both inequalities~\eqref{kurtmth} and~\eqref{skewmth} can be violated in Markovian systems in the presence of thermodynamic forces other than temperature differences, for example chemical potentials. In such a case the detailed balance condition [Eq.~\eqref{detbal}] takes a modified form
\begin{align} \label{detbalmod}
	\frac{k_{ij}^\alpha}{k_{ji}^\alpha}=e^{\beta_\alpha Q_{ij}^\alpha},
\end{align}
where $Q_{ij}^\alpha=E_j-E_i+\mathcal{F}_{ij}^\alpha$ is the heat delivered to the bath $\alpha$ due to transition $j \rightarrow i$ induced by the bath $\alpha$, with $\mathcal{F}_{ij}^\alpha$ being an additional thermodynamic force. While a clear explanation of this phenomenon is lacking, this may be related to the fact that for thermally driven systems the excitation rate to a higher-energy state is always lower than the relaxation rate, i.e., $k_{ij}<k_{ji}$ for $E_i>E_j$. This is no longer true in the presence of other thermodynamic forces since heat increments $Q_{ij}^\alpha$ associated with different baths $\alpha$ may have different signs.

\subsection{Unicyclic Markovian networks} \label{subsec:unmark}
Finally, let me present the bounds obtained for unicyclic Markovian networks, i.e., systems whose states form an ordered chain with transitions allowed only between pairs of neighboring states. Such models can be used, for example, to describe certain biomolecular reactions~\cite{barato2015, barato2015b} or electronic systems~\cite{brandes2008}. The unicyclicity can be mathematically formulated as a condition for the transition rates:
\begin{align}
	\begin{cases}
		k_{ij},k_{ji} \neq 0 \quad & \text{for} \quad i=j + 1 \mod N, \\
		k_{ij},k_{ji} = 0 & \text{otherwise},
	\end{cases}
\end{align}
%
%%%%%%%%%%%%%%%%%%%%%%%%%%%%%%%%%%%%%%%%%%%%%%%%%%%%%%%%%%%%
\begin{figure}
	\centering
	\includegraphics[width=0.7\linewidth]{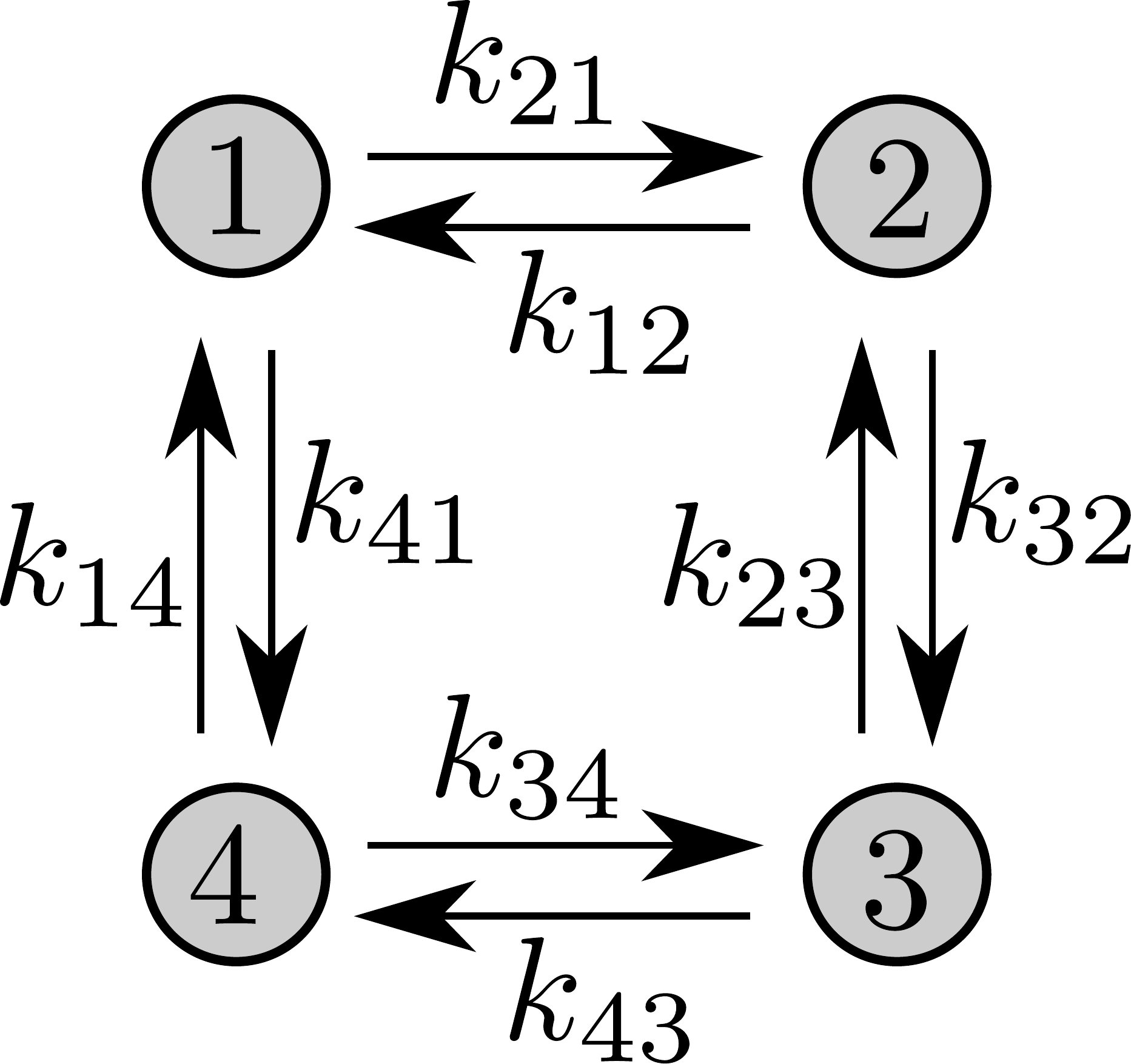}		
	\caption{Schematic representation of a four-state unicyclic Markovian network. Transitions are allowed between pairs of neighboring states such as 1 and 2, 2 and 3 etc., but not between non-neighboring states such as 1 and 3.}
	\label{fig:unischem}
\end{figure}
%%%%%%%%%%%%%%%%%%%%%%%%%%%%%%%%%%%%%%%%%%%%%%%%%%%%%%%%%%%%
%
with $j=1,...,N$. A scheme of an exemplary unicyclic network is presented in Fig.~\ref{fig:unischem}. In this context the relevant quantity are the fluctuations of the winding number, i.e, the number of clockwise rotations around the cycle. It can be defined as
\begin{align}
	\Pi=\Pi_{21}-\Pi_{12},
\end{align}
where $\Pi_{ij}$ is the number of transitions $j \rightarrow i$ within a time interval $[0,t]$. Depending on the system considered, it may correspond to physical observables such as number of biomolecular reactions~\cite{barato2015, barato2015b} or electron jumps in quantum dot systems~\cite{brandes2008}. Scaled cumulants of the winding number can be calculated using the counting-field-dependent generator with elements
\begin{align}
	\begin{cases}
		[W(\lambda)]_{ij}=k_{21} e^\lambda & \text{for} \quad i=2,{ }j=1 \\
		[W(\lambda)]_{ij}=k_{12} e^{-\lambda} & \text{for} \quad i=1,{ }j=2 \\
		[W(\lambda)]_{ij}=W_{ij} & \text{otherwise}.
	\end{cases}
\end{align}
Cumulants can be then calculated using Eqs.~\eqref{calcum1}--\eqref{calcum3}.

Inequalities~\eqref{skewuni}--\eqref{produni} have been obtained using a combination of analytic and numerical methods. First, analytic bounds have been derived by considering unidirectional networks with transitions only in the clockwise direction allowed: 
\begin{align}
	\begin{cases}
		k_{ij} \neq 0 \quad & \text{for} \quad i=j + 1 \mod N, \\
		k_{ij} = 0 & \text{otherwise}.
	\end{cases}
\end{align}
This is described in Sec~\ref{subsec:unid}. Next, their validity to bidirectional networks (with both directions of transitions allowed) has been verified by means of numerical simulations; see Sec.~\ref{subsec:bid} for details.

\subsubsection{Unidirectional networks} \label{subsec:unid}
I will now discuss how the analytic bounds for unidirectional networks have been obtained. First, using numerical optimization techniques implemented within the Wolfram Mathematica environment (functions \mbox{FindMinimum} and \mbox{FindMaximum}) it was inferred that the bounds are always saturated for a specific type of network topology with $k_{21}=ak$ and $k_{ij}=k$ for $j>1$; the parameter $a$ takes different values depending on the considered bound and the number of states. Assuming this type of network topology, the value of $a$ saturating the bounds has been then analytically determined; see Appendix for more details. Specifically, the analytic bound on skewness [Eq.~\eqref{skewuni}] has been found for an arbitrary number of states $N$:
\begin{align} \label{skewunidet}
		\begin{cases}
			\min(\mathcal{S})= -\frac{8-8N+N^2}{16(N-1)^2}& \text{for} \quad a=\frac{1}{N-1}, \\
			\max(\mathcal{S})=1 & \text{for} \quad a \rightarrow 0.
		\end{cases}
	\end{align}
Taking a limit $N \rightarrow \infty$ one gets $\mathcal{S} \geq -1/16$. The other bounds have been derived in the assymptotic limit $N \rightarrow \infty$ in which the saturating value of $a$ scales as $a=A/N$. They read as
\begin{align} \label{kurtunidet}
	&			\begin{cases}
		\min(\mathcal{K})= -\frac{1+\sqrt{5}}{10}& \text{for} \quad A=\frac{3-\sqrt{5}}{2}, \\
		\max(\mathcal{K})=1 & \text{for} \quad A \rightarrow 0,
	\end{cases} \\
	&	\begin{cases}
		\min(\mathcal{S}-\mathcal{K})= 3\frac{107-51 \sqrt{17}}{2048} \approx -0.15 & \text{for} \quad A=\frac{5+\sqrt{17}}{4}, \\
		\max(\mathcal{S}-\mathcal{K})=3\frac{107+51 \sqrt{17}}{2048} \approx 0.465 & \text{for} \quad A =\frac{5-\sqrt{17}}{4},
	\end{cases} \\
	&	\begin{cases}
		\min(\mathcal{S}+ \mathcal{K})= -\frac{8}{27}& \text{for} \quad A=\frac{1}{2}, \\
		\max(\mathcal{S}+\mathcal{K})=2 & \text{for} \quad A \rightarrow 0,
	\end{cases} \\
	&	\begin{cases}
		\min(\mathcal{S} \times \mathcal{K})\approx -0.054 & \text{for} \quad A \approx 0.23, \\
		\max(\mathcal{S} \times \mathcal{K})=1 & \text{for} \quad A \rightarrow 0,
	\end{cases}
\end{align}
where more exactly
\begin{align}
	\min(\mathcal{S} \times \mathcal{K})=\min_A \frac{1-10A+22A^2-12A^3}{(1+A)^8}.
\end{align}
As one may note, the maximum values of skewness and kurtosis $\max \mathcal{S}=\max \mathcal{K}=1$ correspond to the case of $a \rightarrow 0$. In this regime the dynamics of the network is determined by the slowest timescale of the transition $1 \rightarrow 2$, such that the probability of the jump taking place within a short time interval $[t,t+dt]$ is independent of the events occurring in the other time intervals. In such a case the probability of $\Pi$ jumps $1 \rightarrow 2$ taking place within the time window $[0,t]$ is given by the Poisson distribution $P(\Pi)=\langle \Pi \rangle^\Pi e^{-\langle \Pi \rangle}/\Pi!$~\cite{ross1996}, and thus all cumulants are equal to each other.

%
%%%%%%%%%%%%%%%%%%%%%%%%%%%%%%%%%%%%%%%%%%%%%%%%%%%%%%%%%%%%
\begin{figure}
	\centering
	\includegraphics[width=0.90\linewidth]{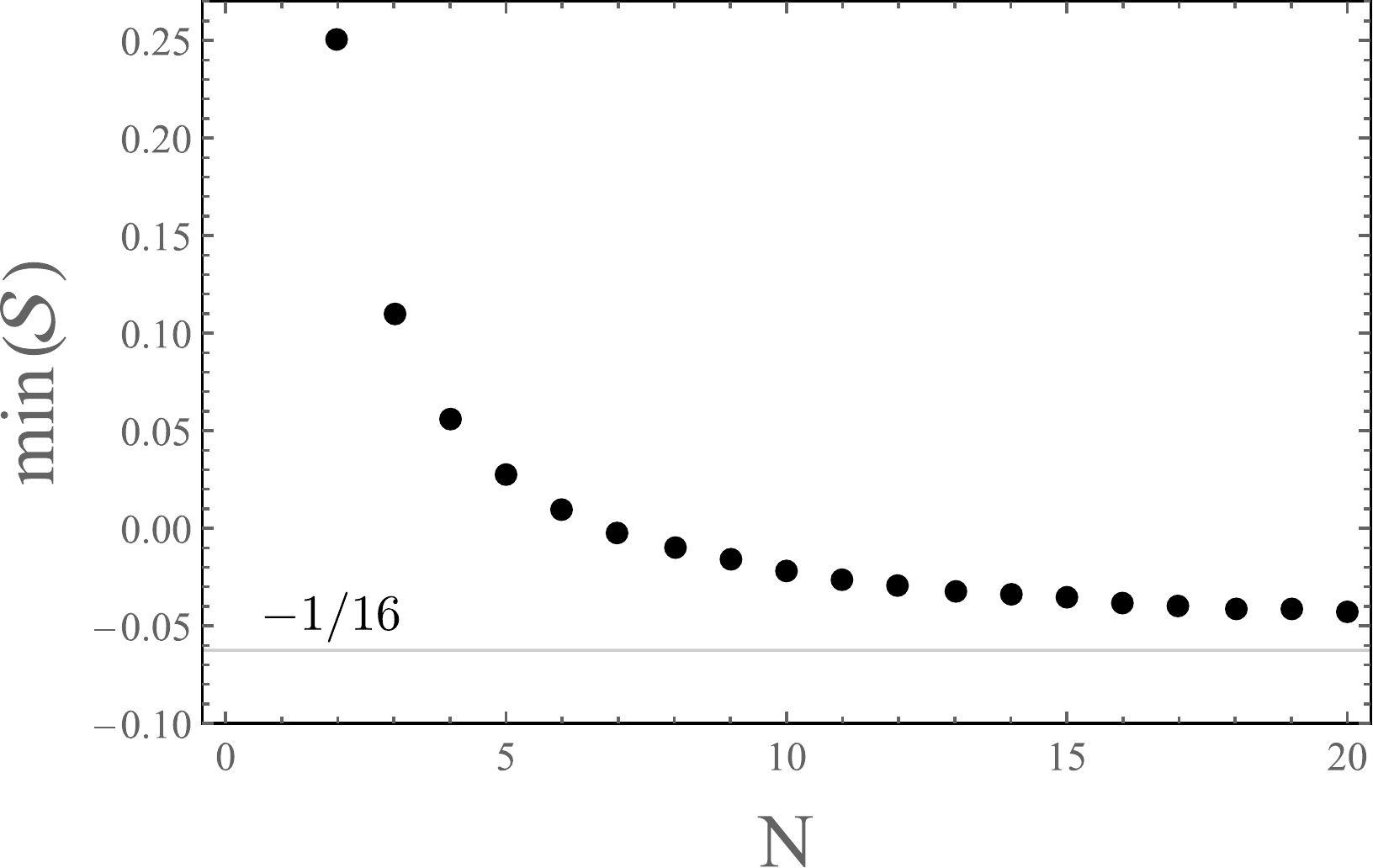}		
	\caption{Minimum value of skewness of the winding number as a function of the number of states $N$.}
	\label{fig:skewuni}
\end{figure}
%%%%%%%%%%%%%%%%%%%%%%%%%%%%%%%%%%%%%%%%%%%%%%%%%%%%%%%%%%%%
%
%
%%%%%%%%%%%%%%%%%%%%%%%%%%%%%%%%%%%%%%%%%%%%%%%%%%%%%%%%%%%%
\begin{figure}
	\centering
	\includegraphics[width=0.90\linewidth]{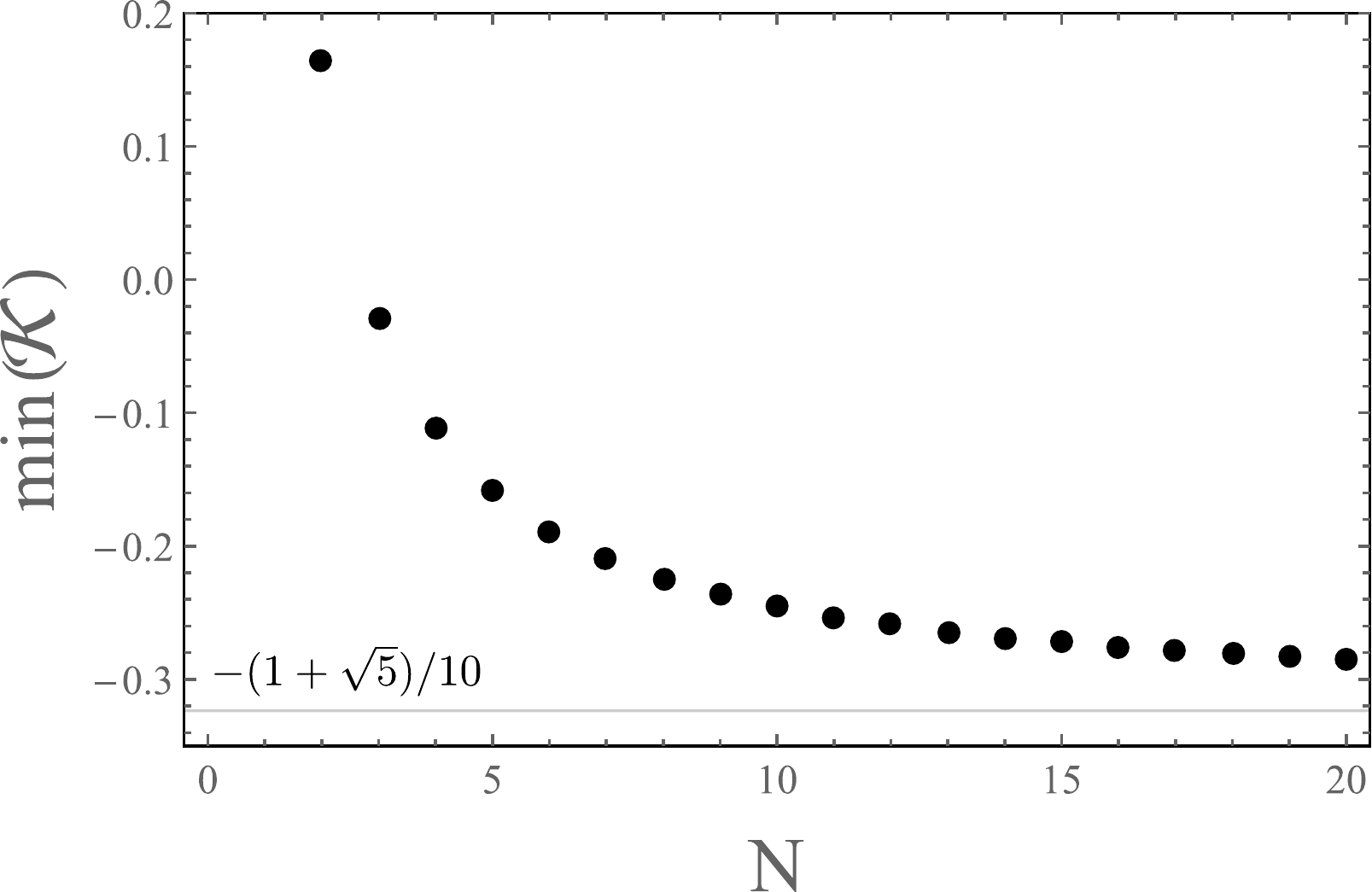}		
	\caption{Minimum value of kurtosis of the winding number as a function of the number of states $N$.}
	\label{fig:kurtuni}
\end{figure}
%%%%%%%%%%%%%%%%%%%%%%%%%%%%%%%%%%%%%%%%%%%%%%%%%%%%%%%%%%%%
%
%
%%%%%%%%%%%%%%%%%%%%%%%%%%%%%%%%%%%%%%%%%%%%%%%%%%%%%%%%%%%%
\begin{figure}
	\centering
	\includegraphics[width=0.85\linewidth]{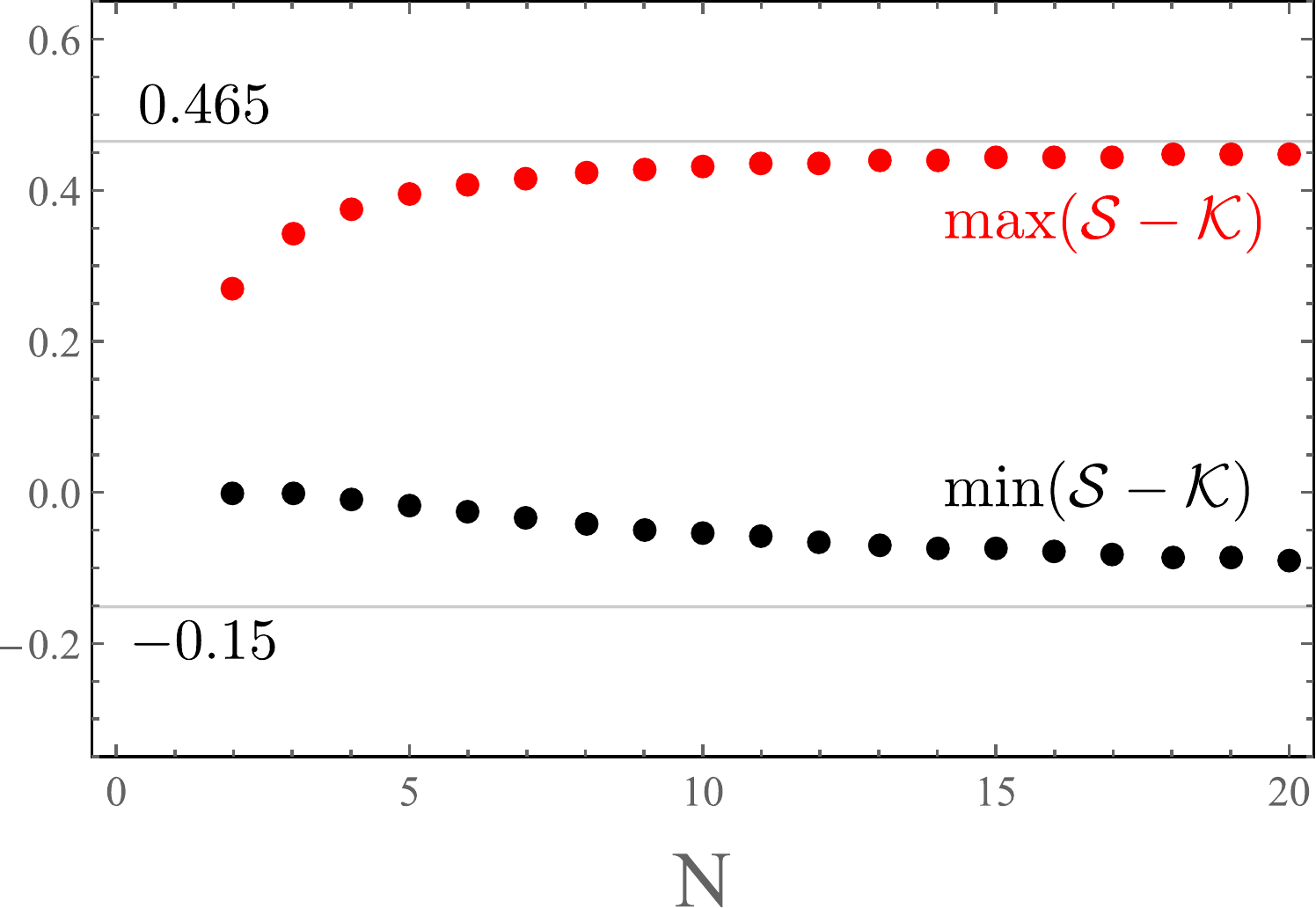}		
	\caption{Minimum (black dots) and maximum (red dots) value of $\mathcal{S}-\mathcal{K}$ of the winding number as a function of the number of states $N$.}
	\label{fig:difuni}
\end{figure}
%%%%%%%%%%%%%%%%%%%%%%%%%%%%%%%%%%%%%%%%%%%%%%%%%%%%%%%%%%%%
%
%
%%%%%%%%%%%%%%%%%%%%%%%%%%%%%%%%%%%%%%%%%%%%%%%%%%%%%%%%%%%%
\begin{figure}
	\centering
	\includegraphics[width=0.90\linewidth]{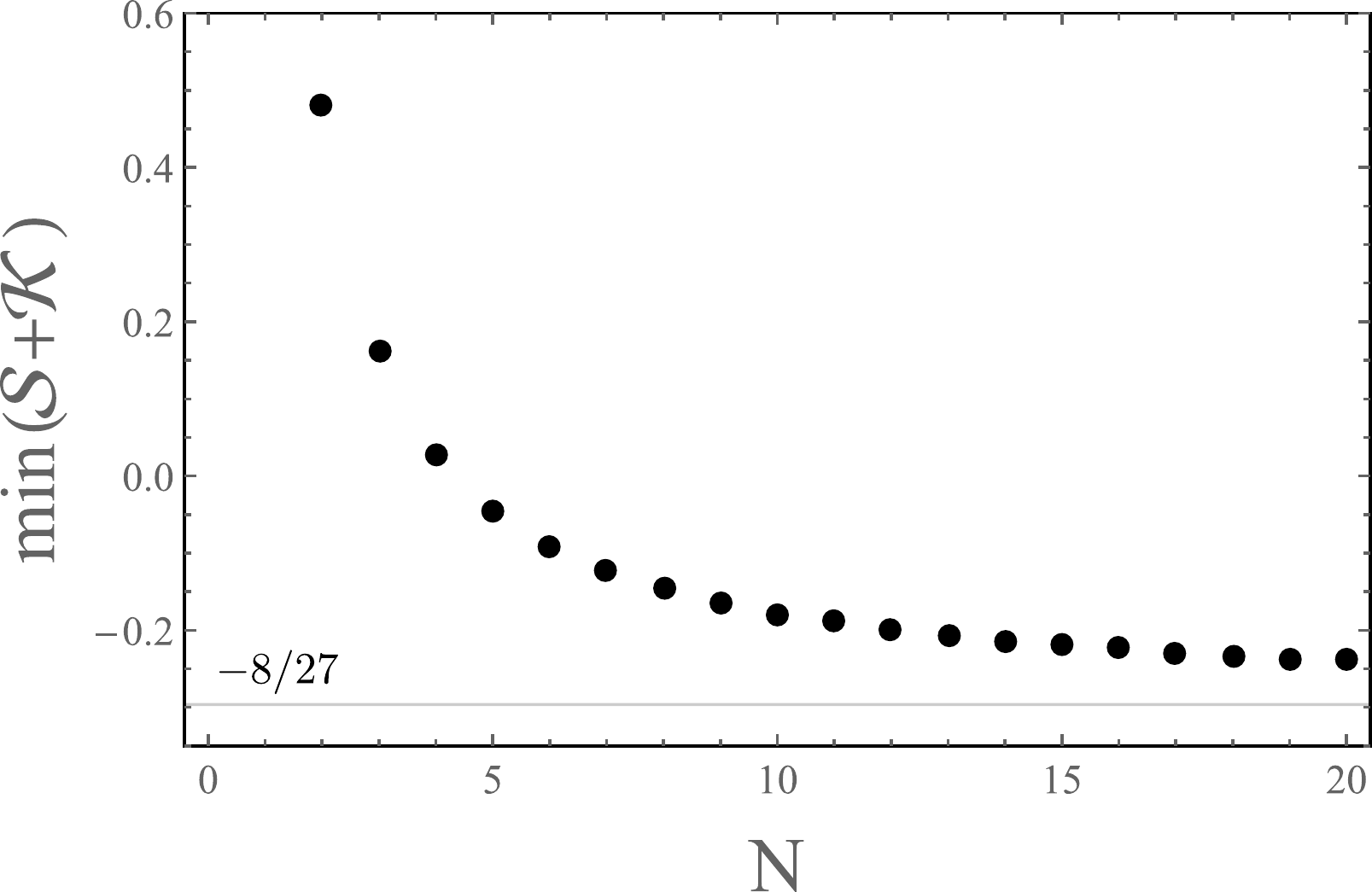}		
	\caption{Minimum value of $\mathcal{S}+\mathcal{K}$ of the winding number as a function of the number of states $N$.}
	\label{fig:sumuni}
\end{figure}
%%%%%%%%%%%%%%%%%%%%%%%%%%%%%%%%%%%%%%%%%%%%%%%%%%%%%%%%%%%%
%
%
%%%%%%%%%%%%%%%%%%%%%%%%%%%%%%%%%%%%%%%%%%%%%%%%%%%%%%%%%%%%
\begin{figure}
	\centering
	\includegraphics[width=0.90\linewidth]{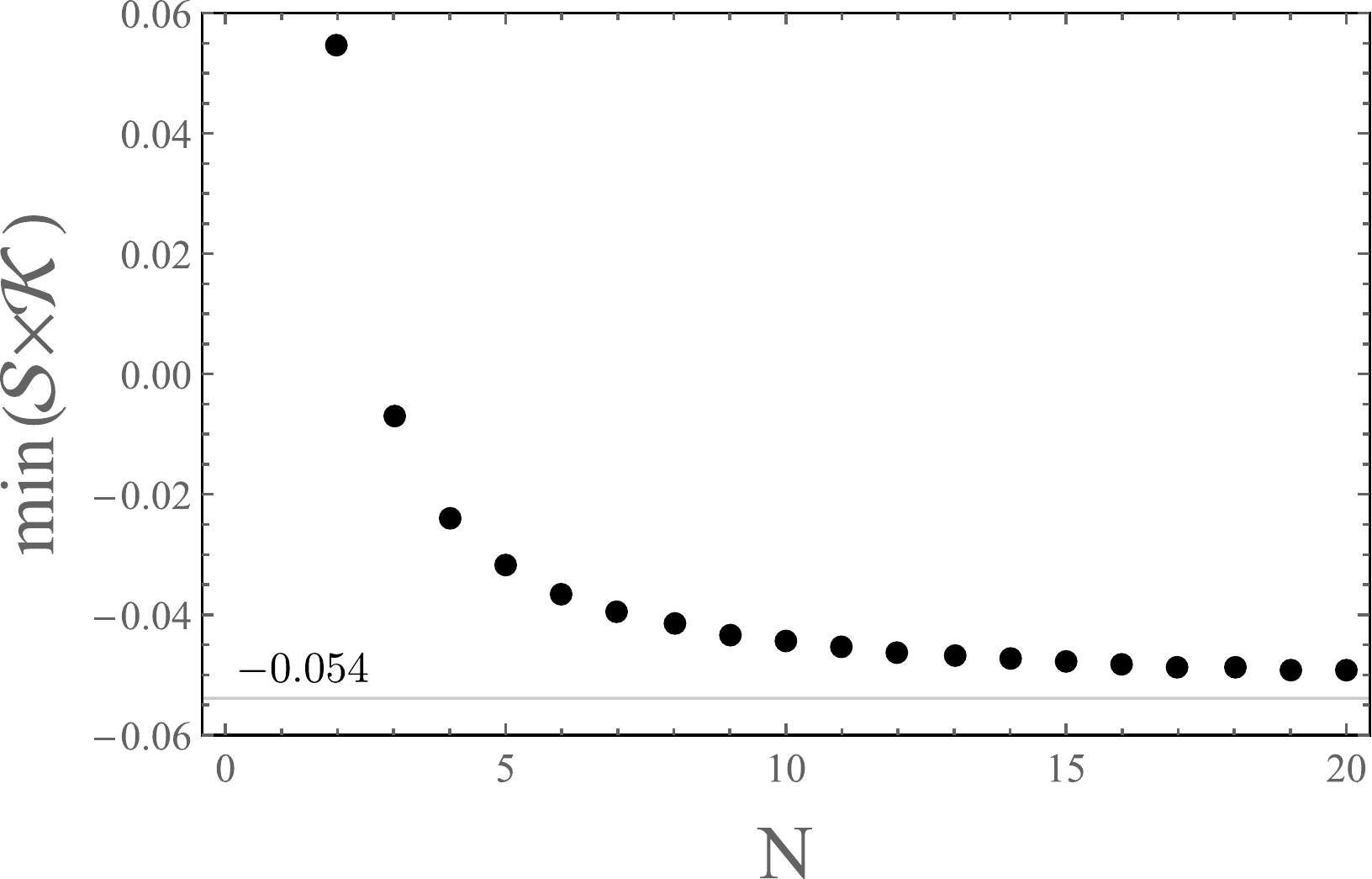}		
	\caption{Minimum value of $\mathcal{S} \times \mathcal{K}$ of the winding number as a function the number of states $N$.}
	\label{fig:produni}
\end{figure}
%%%%%%%%%%%%%%%%%%%%%%%%%%%%%%%%%%%%%%%%%%%%%%%%%%%%%%%%%%%%
%
Though I have focused on the limit $N \rightarrow \infty$, even more tighter bounds can be obtained (at least numerically) for a finite number of states $N$. They are presented in Figs.~\ref{fig:skewuni}--\ref{fig:produni}. Similarly to the previously obtained bounds on the current variance~\cite{koza2002, kolomeisky2007} or fluctuations of waiting times~\cite{moffitt2014, barato2015}, such inequalities can be used to infer a minimum number of states in the unicyclic Markovian network. Furthermore, as demonstrated in the next paragraph, the advantage of the obtained bounds is that they are useful even close to equilibrium, when the variance of the winding number is dominated by the thermal noise and thus relatively insensitive to the network topology.

\subsubsection{Bidirectional networks} \label{subsec:bid}
%
%%%%%%%%%%%%%%%%%%%%%%%%%%%%%%%%%%%%%%%%%%%%%%%%%%%%%%%%%%%%
\begin{figure}
	\centering
	\includegraphics[width=0.90\linewidth]{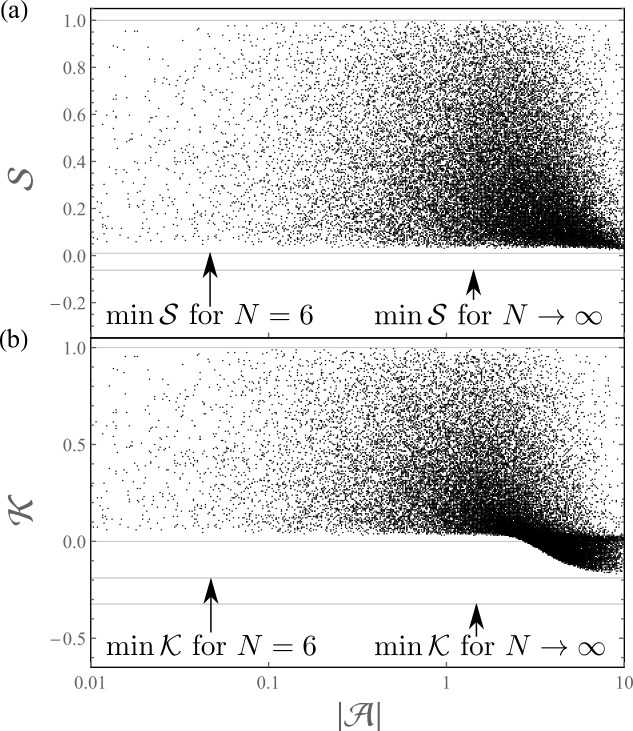}		
	\caption{Skewness and kurtosis of the winding current for 30,000 random unicyclic Markovian networks as a function of the affinity $\mathcal{A}$.}
	\label{fig:bidskew}
\end{figure}
%%%%%%%%%%%%%%%%%%%%%%%%%%%%%%%%%%%%%%%%%%%%%%%%%%%%%%%%%%%%
%
%
%%%%%%%%%%%%%%%%%%%%%%%%%%%%%%%%%%%%%%%%%%%%%%%%%%%%%%%%%%%%
\begin{figure}
	\centering
	\includegraphics[width=0.90\linewidth]{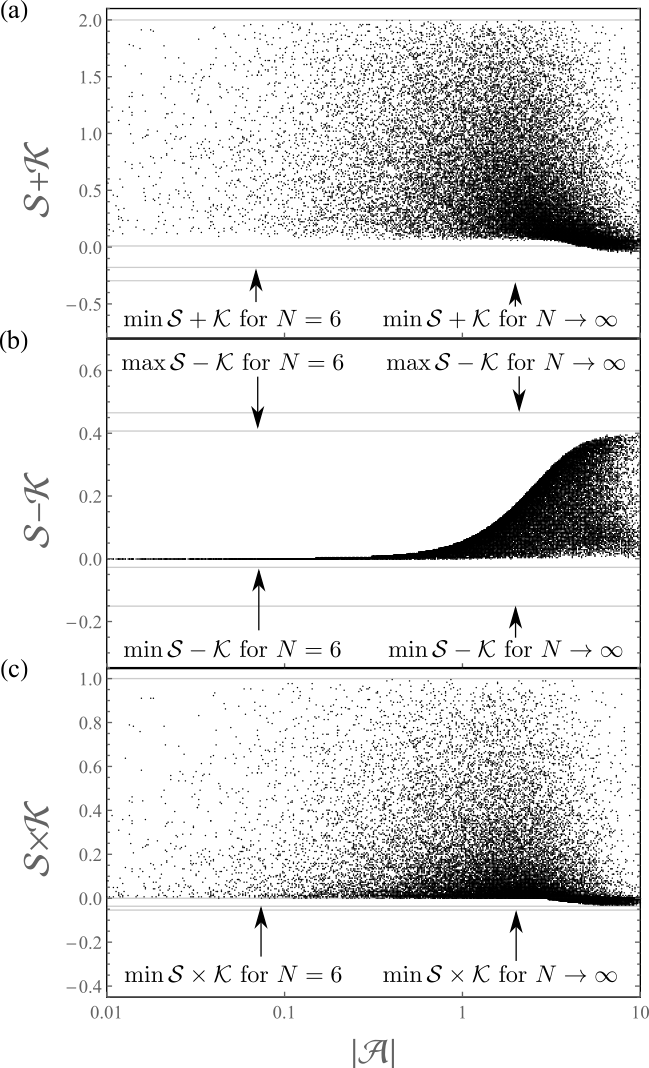}		
	\caption{Sum, difference, and product of skewness and kurtosis of the winding current for 30,000 random unicyclic Markovian networks as a function of the affinity $\mathcal{A}$.}
	\label{fig:bidsum}
\end{figure}
%%%%%%%%%%%%%%%%%%%%%%%%%%%%%%%%%%%%%%%%%%%%%%%%%%%%%%%%%%%%
%

In the next step it was confirmed that the bounds~\eqref{skewuni}--\eqref{produni} are applicable also to bidirectional networks; it has been done by simulating thousands of random Markovian networks with $N \leq 6$; the rates $k_{ij}$ have been taken randomly from the uniform distribution over the interval [0,1]. As an example, Figs.~\ref{fig:bidskew} and~\ref{fig:bidsum} present the simulation outcomes for 30,000 random networks with $N=6$. The results are plotted as a function of the affinity
\begin{align}
	\mathcal{A} = \ln \frac{\prod_{i=1}^N k_{i+1,i}}{\prod_{i=1}^N k_{i,i+1}},
\end{align}
which measures a distance of the network from equilibrium: it is equal to the entropy production (in units of $k_B$) during a single rotation around the cycle in the clockwise direction~\cite{seifert2012}. In particular, the affinity takes a value $\mathcal{A}=0$ at equilibrium while $|\mathcal{A}| \rightarrow \infty$ in the unidirectional case.

As shown in Fig.~\ref{fig:bidskew}, skewness and kurtosis can reach the Poisson limit $\mathcal{S}=\mathcal{K}=1$ for an arbitrary value of the affinity. In contrast, the bound for a minimum value is less tight for a small affinity; this is because skewness and kurtosis can take only nonnegative values close to equilibrium [Eq.~\eqref{eqbound}], while far from equilibrium they can be also negative. A similar behavior is observed for $\mathcal{S}+\mathcal{K}$ and $\mathcal{S} \times \mathcal{K}$ (Fig.~\ref{fig:bidsum}). Interestingly, as shown in Fig.~\ref{fig:bidsum}(b), a most significant dependence on the affinity is observed for the difference ${\mathcal{S}-\mathcal{K}}$ -- it can take values within much wider range far from equilibrium (large $|\mathcal{A}|$) than close to equilibrium (small $|\mathcal{A}|$). Indeed, as implied by Eq.~\eqref{eqbound}, at equilibrium skewness and kurtosis are equal to each other, and thus ${\mathcal{S}-\mathcal{K}=0}$.

\section{Counterexamples} \label{sec:count}
In this sections I will present some exemplary systems in which the obtained bounds can be violated due to going beyond their range of validity. This demonstrates their usefulness for the inference of the underlying physics of the observed transport process.

\subsection{Negativity of skewness and kurtosis of the heat current} \label{subsec:neg}
Before presenting the original research, let me first briefly summarize the relevant results of Saryal \textit{et al}.~\cite{saryal2019} showing how inequalities~\eqref{kurtbos}--\eqref{skewmth} can be broken in thermally driven systems with a unitary component of the dynamics. The authors discussed conditions in which the thermodynamic uncertainty relation~\eqref{unc}, valid for classical Markovian systems, can be violated beyond its range of validity. For a two-terminal setup with bath inverse temperatures $\beta_H$ and $\beta_C$ ($\beta_C>\beta_H$) the thermodynamic uncertainty relation provides a bound on fluctuations of the heat current
\begin{align}
	\frac{\llangle j_h^2 \rrangle}{\llangle j_h^1 \rrangle} \geq \frac{2}{\Delta \beta},
\end{align}
where $\Delta \beta=\beta_C-\beta_H$. Saryal \textit{et al}. shown that in the time-reversal symmetric systems in the linear response regime the relation holds
\begin{align}
\frac{\llangle j_h^2 \rrangle}{\llangle j_h^1 \rrangle}=\frac{2}{\Delta \beta} + \frac{\mathcal{S}^h_\text{lin} }{6} \Delta \beta + \mathcal{O}(\Delta \beta^2),
\end{align}
where $\mathcal{S}^h_\text{lin}$ is the linear-response skewness of the heat current, which is further equal to the equilibrium kurtosis $\mathcal{K}^h_\text{eq}$. This implies that violation of the thermodynamic uncertainty relation close to equilibrium is equivalent to negativity of skewness and kurtosis.

The authors further discussed systems in which the thermodynamic uncertainty relation can be broken close to equilibrium. The first one was a noninteracting fermionic junction. In such a system the cumulants of the heat current can be calculated as
\begin{align}
	\llangle j^n_{h,\gamma \rightarrow \alpha} \rrangle = \int_{-\infty}^\infty \frac{d \omega}{2 \pi} (\omega-\mu_\alpha)^n C_{\alpha\gamma,n}(\omega),
\end{align}
with the functions $C_{\alpha\gamma,n}(\omega)$ defined in Sec.~\ref{subsec:ferm}. Since (as shown in Sec.~\ref{subsec:ferm}) the ratio $C_{\alpha\gamma,3}(\omega)/C_{\alpha\gamma,1}(\omega)$ can be negative (for a high enough transmission function), skewness (and thus kurtosis) of the heat current can also be negative. This is related to coherent, ballistic nature of the electron transport in the high transmission regime, which provides a unitary component of the dynamics. The other model discussed was the spin-boson model attached to baths with structured (non-Ohmic) spectral densities. The reader is referred to Ref.~\cite{saryal2019} for further details.

\subsection{Dynamical channel blockade} \label{subsec:dcb}
Let me now present the original results. First, I will discuss how the obtained bounds can be violated in a classical Markovian model of an interacting electronic system. More specifically, I will focus on a setup consisting of two Coulomb-interacting electronic levels $i \in \{A,B\}$ (for example, spin levels) described by the Hamiltonian:
%%%%%%%%%%%%%%%%%%%%%%%%%%%%%%%%%%%%%%%%%%%%%%%%%%%%%%%%%%%%%%%%%%%%
\begin{figure}
	\centering
	\subfloat[]{\includegraphics[width=0.6\linewidth]{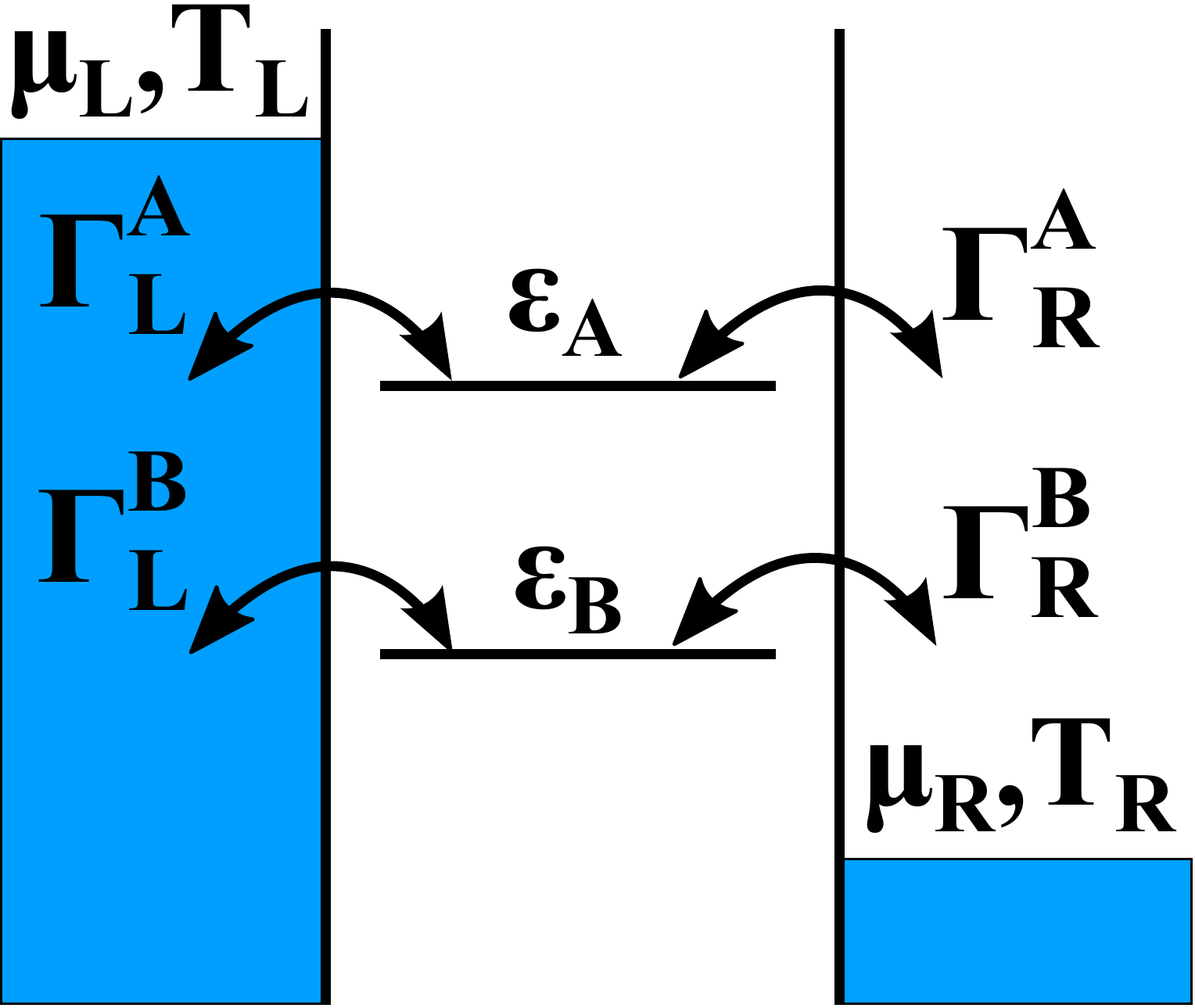}} \\
	\subfloat[]{\includegraphics[width=0.9\linewidth]{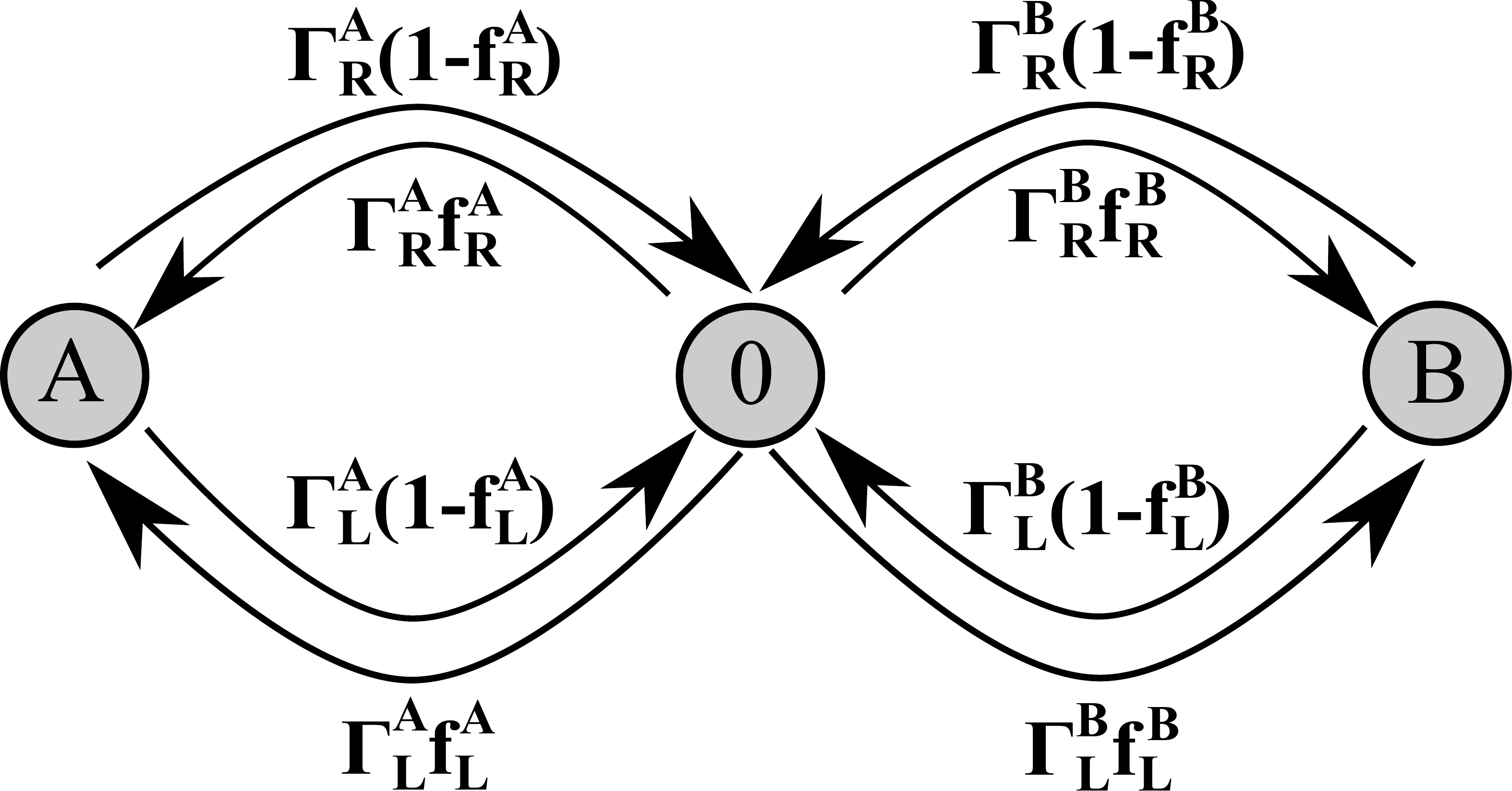}}
	\caption{(a) Scheme of the dynamical channel blockade model. Two electronic levels $A$ and $B$, with energies $\epsilon_A$ and $\epsilon_B$, are coupled to baths $\alpha \in \{L,R\}$ with chemical potentials $\mu_\alpha$ and temperatures $T_\alpha$. $\Gamma_\alpha^{A/B}$ denotes the coupling strength of the bath $\alpha$ to state $A/B$. (b) Three state Markovian model of the system dynamics illustrating its multicyclic nature. Here $f_\alpha^{A/B}=f_\alpha(\epsilon_{A/B})$.}
	\label{fig:dcbscheme}
\end{figure}
%%%%%%%%%%%%%%%%%%%%%%%%%%%%%%%%%%%%%%%%%%%%%%%%%%%%%%%%%%%%%%%%%%%%
\begin{align}
	H_S=\epsilon_A d_A^\dagger d_A + \epsilon_B d_B^\dagger d_B + U d_A^\dagger d_B^\dagger d_A d_B.
\end{align}
Here $d_i^\dagger$ and $d_i$ are the creation and annihilation operators, respectively, $\epsilon_i$ is the level energy and $U$ is the Coulomb interaction strength. The system is coupled to two baths $\alpha \in \{L,R\}$ with temperatures $T_\alpha$ and chemical potentials $\mu_\alpha$; its schematic representation is presented in Fig.~\ref{fig:dcbscheme}~(a). It is further assumed that due to strong Coulomb interaction $U$ only a single occupancy of the dot is allowed; this is referred to as the Coulomb blockade regime. Furthermore, I assume both levels to be unequally coupled to the baths, i.e., $\Gamma_\alpha^A\neq \Gamma_\alpha^B$ where $\Gamma_\alpha^i$ is the coupling strength of the level $i$ to the bath $\alpha$. Such a coupling asymmetry has been demonstrated both theoretically~\cite{bulka2000, belzig2005} and experimentally~\cite{gustavsson2006, ubbelohde2013, hasler2015} to result in the noise enhancement to super-Poissonian values ($\llangle j_{\alpha,p}^2 \rrangle/|\llangle j_{\alpha,p}^1 \rrangle| >1$) in the high voltage regime; this phenomenon has been referred to as the dynamical channel blockade~\cite{belzig2005}.

State of the system is described by the population vector $\mathbf{p}=(p_0,p_A,p_B)$, where $0$ denotes the empty state. The corresponding counting-field-dependent generator takes the form~\cite{bulka2000, belzig2005}
\begin{align} \label{liovdcb} &W^p(\pmb{\lambda})=
	\begin{pmatrix}
		-\Gamma^A_{\text{in},\mathbf{0}}-\Gamma^B_{\text{in},\mathbf{0}} & \Gamma^A_{\text{out},\pmb{\lambda}} & \Gamma^B_{\text{out},\pmb{\lambda}}\\
		\Gamma^A_{\text{in},\pmb{\lambda}} & -\Gamma^A_{\text{out},\mathbf{0}} & 0 \\
		\Gamma^B_{\text{in},\pmb{\lambda}} & 0 & -\Gamma^B_{\text{out},\mathbf{0}}
	\end{pmatrix},
\end{align}
where
\begin{align}
	\Gamma_{\text{in},\pmb{\lambda}}^{i} &=\sum_\alpha \Gamma_\alpha^{i} f_\alpha(\epsilon_{i}) e^{-\lambda_\alpha}, \\
	\Gamma_{\text{out},\pmb{\lambda}}^{i} &= \sum_\alpha \Gamma_\alpha^{i} \left[1-f_\alpha(\epsilon_{i})\right]  e^{\lambda_\alpha}.
\end{align}
Cumulants of the particle current can be then calculated using Eq.~\eqref{calcum1}--\eqref{calcum3}. The coupling strengths to the baths will be parametrized as
\begin{align}
	\Gamma_\alpha^A &=\Gamma_\alpha (1+a_\alpha), \\
	\Gamma_\alpha^B &= \Gamma_\alpha (1-a_\alpha),
\end{align}
where $a_\alpha \in[-1,1]$ is a parameter describing asymmetry of the couplings to the bath $\alpha$. For the sake of simplicity, I will further take $\epsilon_A=\epsilon_B=\epsilon$, $\Gamma_L=\Gamma_R=\Gamma$, $a_L=a_R=a$ and $f_L(\epsilon)=1-f_R(\epsilon)=f$, which holds for $T_L=T_R$ and $\mu_L-\epsilon=\epsilon-\mu_R$. 

%
%%%%%%%%%%%%%%%%%%%%%%%%%%%%%%%%%%%%%%%%%%%%%%%%%%%%%%%%%%%%
\begin{figure}
	\centering
	\includegraphics[width=0.98\linewidth]{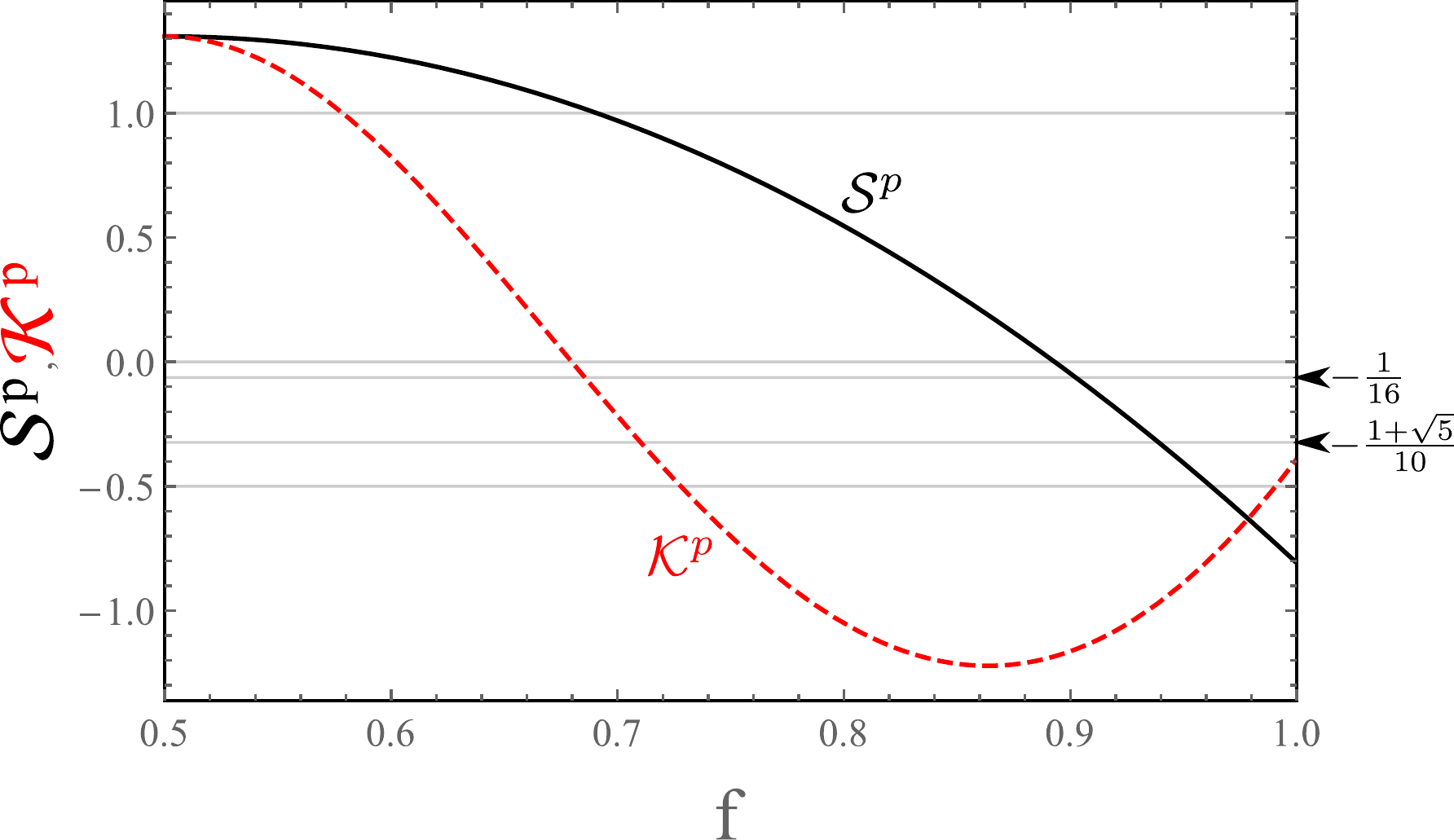}		
	\caption{Skewness (black solid line) and kurtosis (red dashed line) of the particle current in the dynamical channel blockade model with $a=0.65$.}
	\label{fig:skewdcb}
\end{figure}
\begin{figure}
	\centering
	\includegraphics[width=0.9\linewidth]{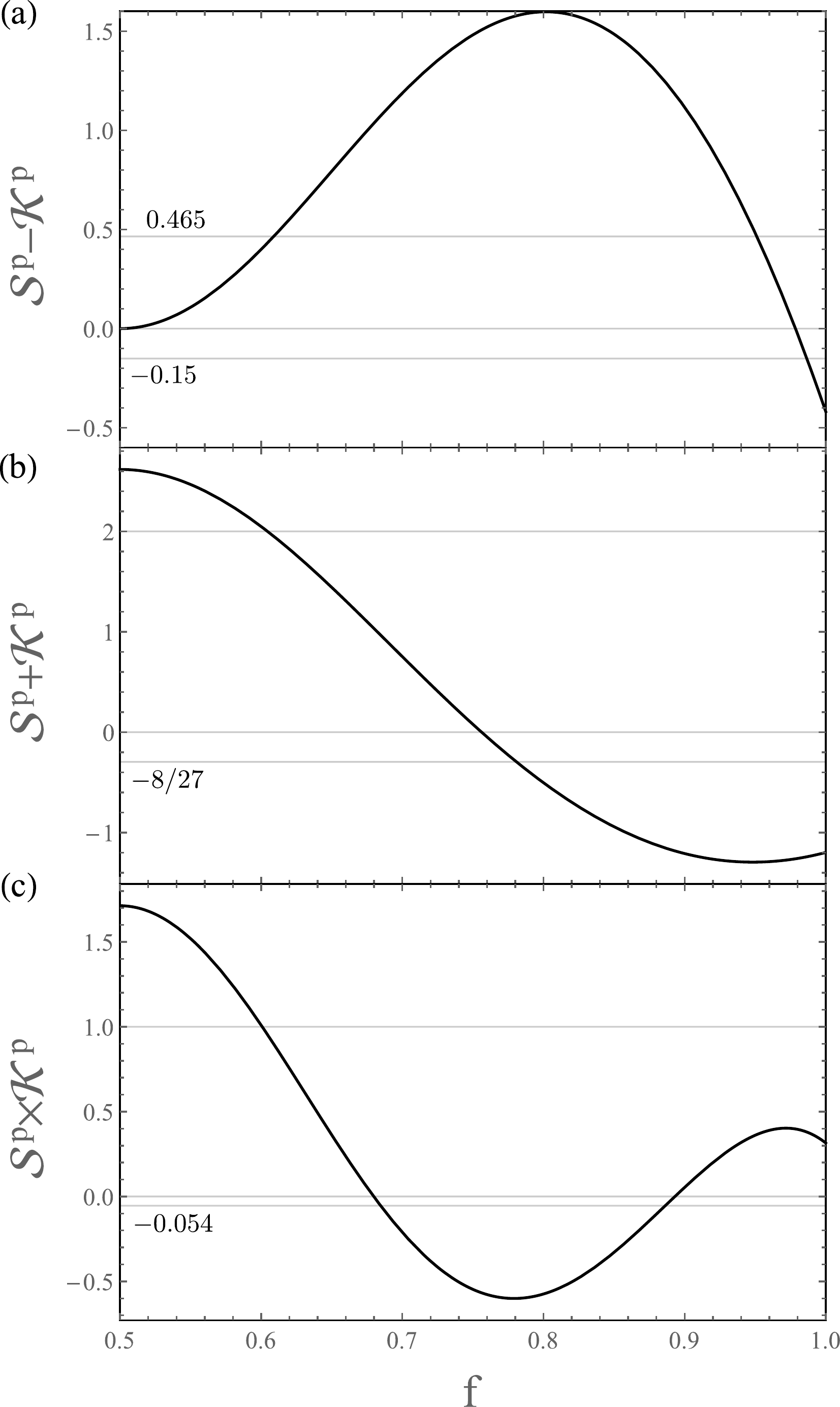}		
	\caption{Demonstration of the violation of bounds~\eqref{difuni}--\eqref{produni} in the dynamical channel blockade model with $a=0.65$.}
	\label{fig:difdcb}
\end{figure}

%%%%%%%%%%%%%%%%%%%%%%%%%%%%%%%%%%%%%%%%%%%%%%%%%%%%%%%%%%%%
%
The calculated skewness and kurtosis of the particle current (which are equal for both baths: $\mathcal{S}^p=\mathcal{S}_L^p=\mathcal{S}_R^p$ etc.) as a function of $f$ are presented in Fig.~\ref{fig:skewdcb}. As one can observe, they can violate bounds $\mathcal{S}^p,{ }\mathcal{K}^p \in [-1/2,1]$ [Eqs.~\eqref{nonintkurt}--\eqref{nonintfskew}], which is related to the interacting nature of the system. The violation of bounds for skewness and kurtosis is observed in slightly different ranges of $f$; therefore, these quantities are complementary indicators of the presence of interactions. Furthermore, one can observe violation of bounds~\eqref{skewuni}--\eqref{kurtuni} derived for unicyclic Markovian networks; as further shown in Fig.~\ref{fig:difdcb} also inequalities~\eqref{difuni}--\eqref{produni} can be broken. This is related to the multicyclic nature of the system, with two different cycles describing transitions $0 \leftrightarrow A$ and $0 \leftrightarrow B$ [see Fig.~\ref{fig:dcbscheme}~(b)]. Therefore, violation of bounds~\eqref{skewuni}--\eqref{produni} can be used to infer the multicyclic nature of the Markovian network underlying the observed transport process.
%
%%%%%%%%%%%%%%%%%%%%%%%%%%%%%%%%%%%%%%%%%%%%%%%%%%%%%%%%%%%%
%%%%%%%%%%%%%%%%%%%%%%%%%%%%%%%%%%%%%%%%%%%%%%%%%%%%%%%%%%%%
%

Let me here note a peculiar merit of the analysis of skewness and kurtosis. Very often, to infer the presence of interactions one analyzes the Fano factor $F=\llangle j^2_p \rrangle/|\llangle j^1_p \rrangle|$, which in noninteracting systems in the high voltage regime takes values $F \in [0,1]$~\cite{blanter2000}. Therefore, $F > 1$ implies the presence of interactions~\cite{bulka2000, belzig2005}. However, this bound is not applicable for small voltages, when the current variance is dominated by the thermal (Johnson-Nyquist) noise and the Fano factor can take arbitrarily large values even in the noninteracting case. In contrast, as Fig.~\ref{fig:skewdcb} implies, violation of bounds~\eqref{nonintkurt}--\eqref{nonintfskew} can be used to infer the presence of interactions even at equilibrium ($f=0.5$).

As Fig.~\ref{fig:skewdcb} further demonstrates, cumulants of the particle current can be negative in the far-from-equilibrium regime. This is also true for the heat current, since for $\epsilon_A=\epsilon_B=\epsilon$ one gets
\begin{align}
	\llangle j_{h,\alpha}^n \rrangle=(\epsilon-\mu_\alpha)^n \llangle j_{p,\alpha}^n \rrangle,
\end{align}
and thus
\begin{align}
	\mathcal{S}_\alpha^h =(\epsilon-\mu_\alpha)^2 \mathcal{S}^p_\alpha, \\
	\mathcal{K}_\alpha^h =(\epsilon-\mu_\alpha)^2 \mathcal{K}^p_\alpha.	
\end{align}
This implies that bounds~\eqref{kurtmth} and~\eqref{skewmth}, stating the nonnegativity of skewness and kurtosis of the heat current in thermally driven two-terminal junctions, are no longer applicable to voltage driven junctions. As follows, violation of Eqs.~\eqref{kurtmth}--\eqref{skewmth} may be used to infer the presence of thermodynamics forces other than temperature differences.

\subsection{Normal metal--superconductor junction} \label{subsec:sup}
In the previous example it was demonstrated that both positive and negative values of skewness and kurtosis violating the bounds $\mathcal{S}^p,{ }\mathcal{K}^p \in [-1/2,1]$ [Eqs.~\eqref{nonintkurt}--\eqref{nonintfskew}] can be observed in classical Markovian systems far from equilibrium. However, as discussed in Sec.~\ref{subsec:neg}, in Markovian networks close to equilibrium only positive values of skewness and kurtosis are allowed. Now I will demonstrate that violation of bounds~\eqref{nonintkurt}--\eqref{nonintfskew} for negative values of skewness and kurtosis (i.e., $\mathcal{S}^p,{ }\mathcal{K}^p \leq -0.5$) can be observed in close-to-equilibrium interacting electronic systems with a unitary component of the dynamics.

The first model considered will be a junction of the normal metal $N$ and the superconductor $S$. I will focus on the wide superconducting gap regime in which quasiparticle (normal electron) tunneling between the normal metal and the superconductor can be neglected. When also electron-electron interactions in the scattering region can be neglected, the fluctuations of the particle current from the normal to the superconducting lead can be described by an analog of the Levitov-Lesovik formula~\cite{muzykantskii1994}
\begin{align} \nonumber
	\chi^p(\lambda)=&\int_{-\infty}^\infty \frac{d \omega}{2 \pi} \ln \left \{1+\mathcal{T}_A(\omega) \left [\left (e^{2\lambda}-1 \right) f_N (\omega) f_N(-\omega) \right. \right. \\ 
	& \left. \left.  +\left (e^{-2\lambda}-1 \right) g_N (\omega) g_N (-\omega) \right] \right \},
\end{align}
where $\mathcal{T}_A(\omega)$ is the transmission function of Andreev tunneling (i.e., a conversion of a Cooper pair from the superconductor into two electrons in the normal metal, or conversely) and, as before, $g_N(\omega)=1-f_N(\omega)$; the chemical potential of Cooper pairs is here fixed at $\mu_{S}=0$. Note that though the considered model is effectively noninteracting, the superconducting electron pairing is itself induced by the electron-electron interactions in the underlying physical system.

Following the steps described in Sec.~\ref{subsec:ferm}, it can be found that the maximum value of skewness and kurtosis 
\begin{align}
\max(\mathcal{S}^p)=\max(\mathcal{K}^p)=4
\end{align} 
is observed in the tunnel junction regime $\mathcal{T}_A(\omega) \rightarrow 0$. It can be noted that in comparison with the noninteracting systems the maximum value is 4 times higher. This corresponds to independent Poissonian tunneling of Cooper pairs. In such a case all particle current cumulants scale as $\llangle j_p^n \rrangle=2^n \llangle j_p^1 \rrangle$ instead of $\llangle j_p^n \rrangle=\llangle j_p^1 \rrangle$ as for the noninteracting, unpaired electrons. The minimum value
\begin{align}
	\min(\mathcal{S}^p)=\min(\mathcal{K}^p)=-2
\end{align}
is, on the other hand, observed in the equilibrium case of $\mu_N=0$ for a boxcar-shaped transmission function
\begin{align}
	\begin{cases}
	\mathcal{T}_A(\omega)=1 & \text{for} \quad -D/2 \leq \omega \leq D/2, \\
	\mathcal{T}_A(\omega)=1 & \text{otherwise},
\end{cases}
\end{align}
in the limit of $D \rightarrow 0$. Again, the value -2 corresponds to a minimum value -0.5, obtained for noninteracting electrons, multiplied by 4 due to electron pairing. Therefore, for an effectively noninteracting normal metal--superconductor junction inequalities~\eqref{nonintkurt}--\eqref{nonintfskew} are replaced by a less tight bound
\begin{align} \label{boundsup}
	\mathcal{S}^p,{ }\mathcal{K}^p \in \left[-2,4 \right].
\end{align}

%
%%%%%%%%%%%%%%%%%%%%%%%%%%%%%%%%%%%%%%%%%%%%%%%%%%%%%%%%%%%%
\begin{figure}
	\centering
	\includegraphics[width=0.75\linewidth]{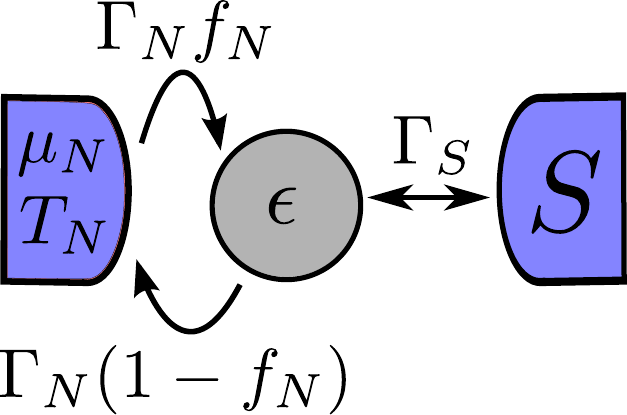}		
	\caption{Scheme of a spin-degenerate quantum dot coupled to the normal lead $N$ (with temperature $T_N$ and chemical potential $\mu_N$) and the superconducting lead $S$.}
	\label{fig:scschem}
\end{figure}
%%%%%%%%%%%%%%%%%%%%%%%%%%%%%%%%%%%%%%%%%%%%%%%%%%%%%%%%%%%%
%
%
%%%%%%%%%%%%%%%%%%%%%%%%%%%%%%%%%%%%%%%%%%%%%%%%%%%%%%%%%%%%
\begin{figure}
	\centering
	\includegraphics[width=0.90\linewidth]{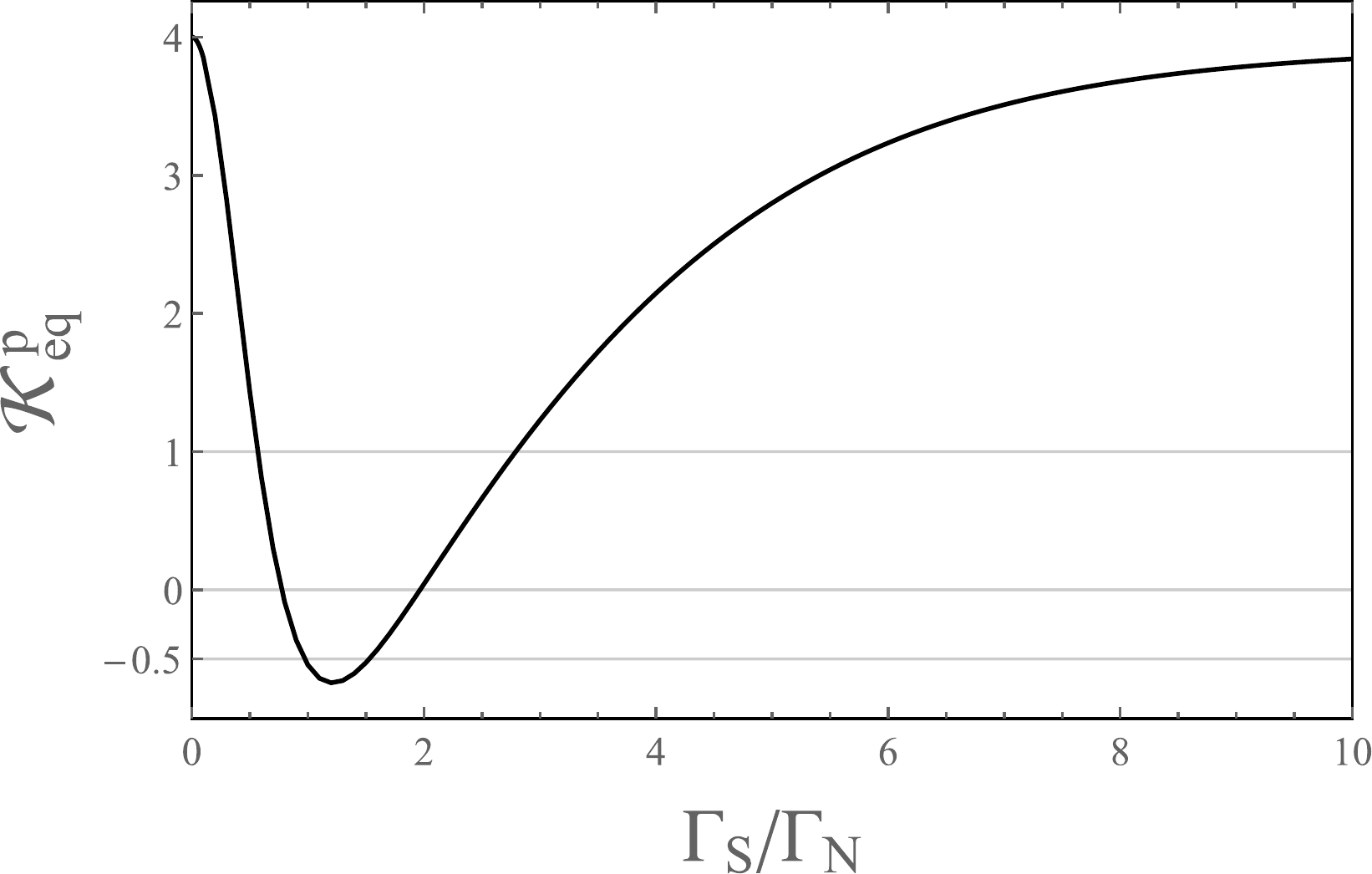}		
	\caption{The equilibrium kurtosis for the normal metal--quantum dot--superconductor junction with $\epsilon=\mu_N=0$ and $k_B T_N=\Gamma_N$.}
	\label{fig:andreev}
\end{figure}
%%%%%%%%%%%%%%%%%%%%%%%%%%%%%%%%%%%%%%%%%%%%%%%%%%%%%%%%%%%%
%
Let us now consider a physically relevant case when the normal and the superconducting lead are coupled through a quantum dot. Such setups have been widely studied both theoretically and experimentally (see review articles~\cite{franceshi2010, rodero2011}). For a noninteracting spin-degenerate dot the transmission function for the Andreev tunneling takes the form~\cite{michalek2013, dong2017}
\begin{align} 
	\mathcal{T}_A(\omega)=\frac{\Gamma_N^2 \Gamma_S^2}{4 \text{Abs} \left[(\omega+i\Gamma_N/2)^2-\Gamma_S^2/4 \right]},
\end{align}
where $\Gamma_N$ and $\Gamma_S$ are the coupling strengths to the normal and the superconducting leads, respectively, and the electron level energy $\epsilon=0$ has been taken for the sake of simplicity. Quite notably, the considered model provides a particularly elegant qualitative interpretation of the unitary component of the dynamics which leads to violation of bound~\eqref{eqbound}: It is related to the coherent oscillations of Cooper pairs between the quantum dot and the superconductor~\cite{rajabi2013}. More formally, the quantum dot attached to the superconducting lead can described by the effective Hamiltonian~\cite{rozhkov2000, braggio2011}
\begin{align}
H_\text{eff}=\sum_{\sigma=\uparrow,\downarrow} \epsilon d_\sigma^\dagger d_\sigma+\frac{\Gamma_S}{2} \left(d_\uparrow^\dagger d_\downarrow^\dagger + d_\downarrow d_\uparrow \right),
\end{align}
where the second term describes the coherent oscillations of Cooper pairs with with a frequency $\Gamma_S$/2.

The equilibrium kurtosis of the particle current $\mathcal{K}_\text{eq}^p$ for a quantum dot model as a function of $\Gamma_S/\Gamma_N$ is presented in Fig.~\ref{fig:andreev}. As one can observe, it reaches a tunnel junction limit $\mathcal{K}_\text{eq}^p=4$ for an asymmetric coupling $\Gamma_S/\Gamma_N \rightarrow 0$ or $\Gamma_S/\Gamma_N \rightarrow \infty$; it can be here noted that for $\Gamma_S \gg \Gamma_N$ the system can be effectively described by an effectively classical Markovian master equation~\cite{braggio2011}. Most importantly, kurtosis can be reduced below -0.5 for $\Gamma_S \approx \Gamma_N$, which -- as discussed before -- implies both the presence of interactions [violation of bound~\eqref{nonintkurt}] and of the unitary component of the dynamics [violation of bound~\eqref{eqbound}]. The second fact can be understood as follows: for $\Gamma_S \approx \Gamma_N$ the tunneling rate to the normal lead $\Gamma_N$ is of the same order of magnitude as the frequency of coherent oscillations of Cooper pairs $\Gamma_S/2$, which makes the classical Markovian description [providing the validity of bound~\eqref{eqbound}] no longer applicable.

Finally, it can be noted that values of skewness and kurtosis violating inequalities~\eqref{nonintkurt}--\eqref{nonintfskew} have been observed in the strongly correlated quantum dot in which the electron pairing was a result of the Kondo effect rather than superconducting correlations~\cite{komnik2005}. Furthermore, also a less tight bound~\eqref{boundsup} can be violated in transport between two superconducting leads dominated by multiple Andreev reflections~\cite{cuevas2003}.

\subsection{Triple quantum dot} \label{subsec:tqd}
%
%%%%%%%%%%%%%%%%%%%%%%%%%%%%%%%%%%%%%%%%%%%%%%%%%%%%%%%%%%%%
\begin{figure}
	\centering
	\includegraphics[width=0.95\linewidth]{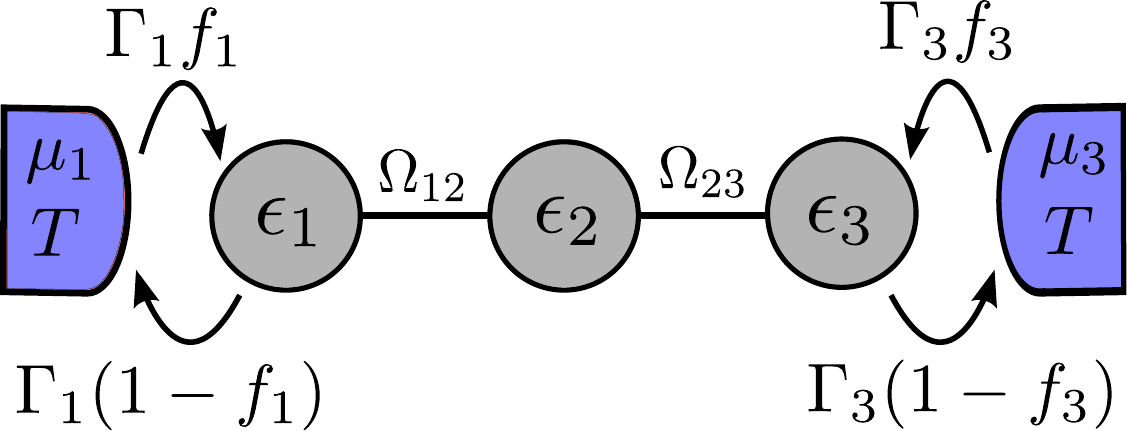}		
	\caption{Scheme of the triple quantum dot molecule connected to two leads 1 and 3 with the same temperature $T$ and chemical potentials $\mu_1$ and $\mu_3$.}
	\label{fig:tqdschem}
\end{figure}
%%%%%%%%%%%%%%%%%%%%%%%%%%%%%%%%%%%%%%%%%%%%%%%%%%%%%%%%%%%%
%
Finally, it will be demonstrated that the equilibrium kurtosis $\mathcal{K}^p_\text{eq}$ can be reduced below -0.5 in interacting, quantum coherent electronic systems without superconducting pairing. In particular, I will consider a triple quantum dot coupled in series described by the Hamiltonian
\begin{align} \nonumber
	H_S&=\sum_{i=1}^3 \epsilon_i d^\dagger_i d_i + \sum_{i=1,j>i}^3 U_{ij} d^\dagger_i d^\dagger_j d_i d_j \\ &+\Omega_{12} (d^\dagger_1 d_2+d_2^\dagger d_1)+\Omega_{23} (d^\dagger_2 d_3+d_3^\dagger d_2),
\end{align}
where $U_{ij}$ and $\Omega_{ij}$ are the Coulomb interaction and the tunnel coupling between the quantum dots, respectively. Scheme of the system is presented in Fig.~\ref{fig:tqdschem}. A strong Coulomb interaction $U_{ij} \rightarrow \infty$ will be further assumed such that only a zero or a single occupancy of the molecule is allowed. The system can be then described by the effective Hamiltonian
\begin{align} \nonumber
	H_\text{eff} = \sum_{i=1}^3 \epsilon_i |i \rangle \langle i|&+\Omega_{12} (|1\rangle \langle 2|+|2\rangle \langle 1|) \\ &+\Omega_{23} (|2\rangle \langle 3|+|3\rangle \langle 2|),
\end{align}
where $|i \rangle$ denotes the occupied state of the $i$th dot and $|0\rangle$ denotes the empty state. The dots 1 and 3 are connected to the baths 1 and 3 with the same temperature $T$. For $\Gamma_i,{ }\Omega_{ij} \ll k_B T$ dynamics of the system can be approximately well described by a local master equation in the Lindblad form~\cite{trushechkin2016, hofer2017}
\begin{align} \nonumber
\dot{\rho}=&-i \left[H_\text{eff},\rho \right]+\sum_{i=1,3} \Gamma_i f_i(\epsilon_i) \left(L_i \rho L_i^\dagger- \frac{1}{2} \left \{L_i^\dagger L_i,\rho \right \} \right) \\ 
&+\sum_{i=1,3} \Gamma_i \left[1-f_i(\epsilon_i) \right] \left(L_i^\dagger \rho L_i - \frac{1}{2} \left \{L_i L_i^\dagger,\rho \right \} \right),
\end{align}
where $\rho$ is the density matrix of the system, $\Gamma_i$ is the coupling strength of the dot $i$ to the bath $i$ and $L_i=|i \rangle \langle 0|$ is the jump operator; it should be here noted that local master equations of such type may provide certain unphysical results beyond their range of validity and therefore should be applied with care~\cite{carmichael1973, levy2014, stockburger2016}. For the sake of simplicity, here I do not treat the level renormalization induced by the Coulomb interaction~\cite{wunsch2005, konig2003, braun2004} explicitly, but rather take the energies $\epsilon_i$ entering $H_\text{eff}$ to be the renormalized values.

As in the classical Markovian systems, current cumulants can be calculated using the counting-field-dependent generator $W^p(\pmb{\lambda})$. It can be defined using the Liouville space representation in which the $N \times N$ density matrix $\rho$ is expressed as $N^2$ row vector $\tilde{\rho}$ such that element $\rho_{ij}$ of the density matrix corresponds to $(i-1)N+j$ element of the vector $\tilde{\rho}$~\cite{machnes2014, am-shallem2015, uzdin2016}. The generator takes the form
\begin{widetext}
\begin{align} \nonumber
	W^p(\pmb{\lambda}) &=-i \left( \mathds{1}_4 \otimes H_\text{eff}-H_\text{eff}^T \otimes \mathds{1}_4 \right) +\sum_{i=1,3} \Gamma_i f_i(\epsilon_i) \left[\left(L_i^\dagger \right)^T L_i e^{-\lambda_i} - \frac{1}{2}\mathds{1}_4 \otimes L_i^\dagger L_i - \frac{1}{2} \left(L_i^\dagger L_i \right)^T \otimes \mathds{1}_4 \right] \\
	&+ \sum_{i=1,3} \Gamma_i  \left[1-f_i(\epsilon_i) \right] \left[L_i^T L_i^\dagger e^{\lambda_i} - \frac{1}{2}\mathds{1}_4 \otimes L_i L_i^\dagger - \frac{1}{2} \left(L_i L_i^\dagger \right)^T \otimes \mathds{1}_4 \right],
\end{align}
\end{widetext}
where the counting fields were introduced following the procedure presented in Ref.~\cite{bruderer2014}; here $\mathds{1}_4$ is $4 \times 4$ identity matrix while $H_\text{eff}$ and $L_i$ are represented in the matrix form in the basis $\{|0\rangle, |1 \rangle, |2 \rangle, |3 \rangle\}$ as
\begingroup
\allowdisplaybreaks
\begin{align}
	H_\text{eff} &=\text{diag} (0,\epsilon_1,\epsilon_2,\epsilon_3), \\
	L_1 & = 	\begin{pmatrix}
		0 & 0 & 0 & 0 \\
		1 & 0 & 0 & 0 \\
		0 & 0 & 0 & 0 \\
		0 & 0 & 0 & 0
	\end{pmatrix}, \\
	L_3 & = 	\begin{pmatrix}
		0 & 0 & 0 & 0 \\
		0 & 0 & 0 & 0 \\
		0 & 0 & 0 & 0 \\
		1 & 0 & 0 & 0
	\end{pmatrix}.
\end{align}
\endgroup

%
%%%%%%%%%%%%%%%%%%%%%%%%%%%%%%%%%%%%%%%%%%%%%%%%%%%%%%%%%%%%
\begin{figure}
	\centering
	\includegraphics[width=0.90\linewidth]{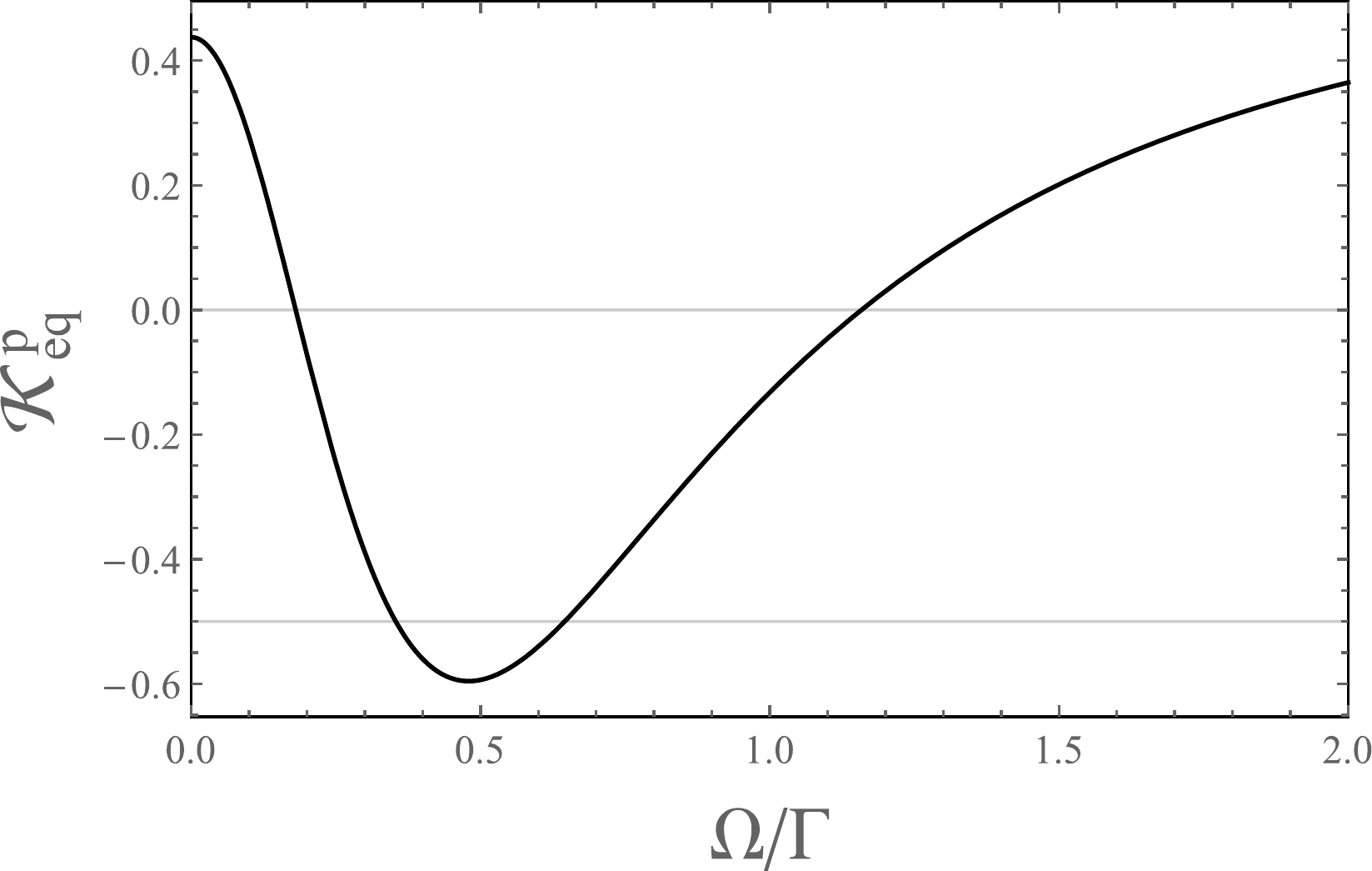}		
	\caption{The equilibrium kurtosis for the triple quantum dot molecule with $\epsilon_1=\epsilon_2=\epsilon_3=\mu_1=\mu_2$, $\Omega_{12}=\Omega_{23}=\Omega$, $\Gamma_1=\Gamma_3=\Gamma$ and $\Gamma,{ }\Omega \ll k_B T$}
	\label{fig:tqd}
\end{figure}
%%%%%%%%%%%%%%%%%%%%%%%%%%%%%%%%%%%%%%%%%%%%%%%%%%%%%%%%%%%%
%
The particle current cumulants can be then calculated using Eqs.~\eqref{calcum1}--\eqref{calcum3}. Taking $\epsilon_1=\epsilon_2=\epsilon_3=\mu_1=\mu_2$, $\Omega_{12}=\Omega_{23}=\Omega$ and $\Gamma_1=\Gamma_3=\Gamma$ one gets
\begin{align}
	\mathcal{K}_\text{eq}^p=\frac{7 \Gamma^4-220 \Gamma^2 \Omega^2+160 \Omega^4}{16(\Gamma^2+4 \Omega^2)^2},
\end{align}
where the kurtosis is equal for both baths: $\mathcal{K}_\text{eq}^p=\mathcal{K}_{1,\text{eq}}^p=\mathcal{K}_{3,\text{eq}}^p$. The results are presented in Fig.~\ref{fig:tqd}. As one can observe, kurtosis can go below -0.5 at equilibrium, thus simultaneously violating both the bound~\eqref{nonintkurt} (which indicates the presence of interactions) and the bound~\eqref{eqbound} (which indicates the presence of the unitary dynamics); specifically, it reaches a minimum value $\min(\mathcal{K}_\text{eq}^p) \approx -0.6$ for $\Omega/\Gamma \approx 0.48$. As noted in Sec.~\ref{subsec:neg}, negativity of the equilibrium kurtosis is related to the violation of the classical thermodynamic uncertainty relation due to presence of coherent electron oscillations, which has been already observed for a simpler double-dot system~\cite{ptaszynski2018, agarwalla2018}; however, though negative $\mathcal{K}_\text{eq}^p$ can be already observed for the double-dot setup, the reduction of $\mathcal{K}_\text{eq}^p$ below -0.5 (which indicates the presence of interactions) requires the triple-dot system.

\section{Conclusions} \label{sec:concl}
In conclusion, the paper presents bounds on skewness and kurtosis of steady state currents applicable to several classes of physical systems. The obtained inequalities have been either analytically derived or numerically conjectured. It was also demonstrated how these bounds can be broken by going beyond their range of applicability, which provides information about the underlying physics of the observed transport setup.

Most importantly, the main value of the obtained bounds results from their complementarity: the measurement of a single quantity (skewness or kurtosis) can provide a broad spectrum of information about the dynamics and thermodynamics of the system. For example, in the dynamical channel blockade system analyzed in Sec.~\ref{subsec:dcb} skewness and kurtosis provide information about three independent facts: the presence of interactions [violation of bounds~\eqref{nonintkurt}--\eqref{nonintfskew}], multicyclic nature of the Markovian network [violation of bounds~\eqref{skewuni}--\eqref{produni}], and the presence of thermodynamic forces other than temperature differences [violation of bounds~\eqref{kurtmth}--\eqref{skewmth}]. Analogously, as shown in Secs.~\ref{subsec:sup} and~\ref{subsec:tqd}, kurtosis of the particle current going below $-0.5$ at equilibrium implies not only the presence of interactions [violation of bound~\eqref{nonintkurt}], but also of a unitary component of the dynamics [violation of bound~\eqref{eqbound}].

Furthermore, the presented inequalities have peculiar advantages in comparison with the already known bounds on the current noise. As previously demonstrated, the bounds on the current variance can be used to detect the presence of interactions in fermionic systems~\cite{blanter2000, bulka2000, belzig2005} or infer the minimum number of states in the Markovian network~\cite{koza2002, kolomeisky2007} in far-from-equilibrium conditions. However, these inequalities cease to be useful close to equilibrium, when the current variance is dominated by the thermal noise. The obtained bounds on skewness and kurtosis, instead, are applicable for these purposes arbitrarily close to equilibrium (see Secs.~\ref{subsec:unmark} and~\ref{subsec:dcb}); indeed, a similar insensitivity of skewness to the thermal noise has been previously reported by Levitov and Reznikov~\cite{levitov2004}. This highlights the merits of the analysis of higher-order cumulants for the characterization of transport processes.

\begin{acknowledgments}
	The author has been supported by the National Science Centre, Poland, under the project No.~2017/27/N/ST3/01604, and by the Scholarships of Minister of Science and Higher Education.
\end{acknowledgments}

\appendix*

\section{Derivation of bounds \eqref{skewuni}--\eqref{produni}}
This Appendix discusses the details of derivation of bounds \eqref{skewuni}--\eqref{produni} for unidirectional unicyclic networks. To this end the analytic formulas for scaled cumulants of the winding number have been used. While they can be obtained directly using Eqs.~\eqref{calcum1}--\eqref{calcum3}, it is more convenient to express them as~\cite{albert2011}
\begin{align}
	\llangle j^1 \rrangle & = \frac{1}{\kappa_1}, \\
	\llangle j^2 \rrangle & = \frac{\kappa_2}{\kappa_1^3}, \\
	\llangle j^3 \rrangle & = 3\frac{\kappa_2^2}{\kappa_1^5}-\frac{\kappa_3}{\kappa_1^4}, \\
	\llangle j^4 \rrangle & = 15\frac{\kappa_2^3}{\kappa_1^7}-10 \frac{\kappa_2 \kappa_3}{\kappa_1^6}+\frac{\kappa_4}{\kappa_1^5},
\end{align}
where $\kappa_n$ is the $n$th cumulant of the waiting times between the subsequent jumps $1 \rightarrow 2$; such relations hold for unicyclic networks due to the renewal property (the successive waiting times are uncorrelated). The cumulants $\kappa_n$ can be determined using the Laplace transform of the waiting time distribution $\tilde{w}(s)$~\cite{albert2011},
\begin{align}
	\kappa_n = (-1)^n \frac{\partial^n}{\partial s^n} \ln \tilde{w}(s),
\end{align}
where for unicyclic networks~\cite{brandes2008}
\begin{align}
	\tilde{w}(s)=\prod_{i=1}^N \frac{k_{i+1,i}}{k_{i+1,i}+s},
\end{align}
with $k_{N+1,N}=k_{1N}$. Explicitly
\begin{align}
	\kappa_1 = \sum_{i=1}^N k_{i+1,i}^{-1}, \\
	\kappa_2 = \sum_{i=1}^N k_{i+1,i}^{-2}, \\
	\kappa_3 = 2\sum_{i=1}^N k_{i+1,i}^{-3}, \\
	\kappa_4 = 6\sum_{i=1}^N k_{i+1,i}^{-4}.
\end{align}
As discussed in Sec.~\ref{subsec:unmark}, it was numerically conjectured that the minimum and maximum values of $\mathcal{S}$, $\mathcal{K}$, $\mathcal{S}+\mathcal{K}$, $\mathcal{S}-\mathcal{K}$ and $\mathcal{S} \times \mathcal{K}$ are always obtained for a specific type of network topology, with all rates $k_{ij}$ equal to $k$ apart from one (here $k_{21}$) equal to $ak$, with the parameter $a$ depending on the optimized quantity and the number of states $N$. For such a topology the waiting time cumulants take the form
\begin{align}
	\kappa_1 &= k^{-1} (N-1+a^{-1}), \\
	\kappa_2 &= k^{-2} (N-1+a^{-2}), \\
	\kappa_3 &= 2k^{-3} (N-1+a^{-3}), \\
	\kappa_4 &= 6k^{-4} (N-1+a^{-4}).
\end{align}

Using the equations above skewness and kurtosis can be expressed analytically. The values of $a$ minimizing or maximizing the skewness can be found by solving the equation
\begin{align}
	\frac{\partial \mathcal{S}}{\partial a}=0.
\end{align}
The maximum value $\max (\mathcal{S})=1$ corresponds to the limit of $a \rightarrow 0$ when fluctuations are fully determined by the slowest timescale of the transition $1 \rightarrow 2$. The minimum value $\min (\mathcal{S})$ is found for $a=1/(N-1)$; it is given by Eq.~\eqref{skewunidet}.

Let us now consider the kurtosis. First, as for the skewness, $\max (\mathcal{K})=1$ in the limit of $a \rightarrow 0$. The problem of finding the minimum value is more involved, namely, $\text{argmin}_a(\mathcal{K})$ cannot be found analytically since it is given by the root of a fifth degree polynomial of $a$. However, one may numerically infer that
\begin{itemize}
	\item $\min (\mathcal{K})$ decreases monotonically with $N$ (see Fig.~\ref{fig:kurtuni}),
	\item $\text{argmin}_a(\mathcal{K}) \propto 1/N$ for large $N$.
\end{itemize}
Upon substituting $a \rightarrow A/N$ one gets
\begin{align}
	\lim_{N \rightarrow \infty} \mathcal{K} = \frac{1-8A+6A^2}{(1+A)^4}.
\end{align}
Using the expression above one easily obtains Eq.~\eqref{kurtunidet}. The other bounds have been obtained in a similar manner.

\end{document}